\pdfoutput=1
\documentclass[aps,prc,twocolumn,superscriptaddress,showpacs,floatfix
,nofootinbib]{revtex4-2}
\usepackage{graphicx}
\usepackage{amsfonts}
\usepackage{amsmath}
\usepackage{amssymb}
\usepackage[super]{nth}
\usepackage{bm}
\usepackage[normalem]{ulem}
\usepackage{xcolor}
\usepackage{longtable}
\usepackage{epstopdf}
\usepackage{mathtools}
\usepackage[T1]{fontenc}
\usepackage[utf8]{inputenc}
\usepackage{ae}
\usepackage{aecompl}
\usepackage[activate={true,nocompatibility},final,tracking=true,kerning=true,spacing=true]{microtype}
\usepackage[colorlinks = true, %Colours links instead of stupid boxes
            urlcolor   = blue, %Colour for external hyperlinks
            linkcolor  = blue, %Colour of internal links
            citecolor  = blue,
            linktoc    = all,
            bookmarks  = true,
            bookmarksnumbered = true,
            bookmarksdepth = 4,
            hyperindex,breaklinks]{hyperref}

\begin{document}
\title{Path-integral quantum Monte Carlo calculations of light nuclei}
\author{Rong Chen}
\email{chen.rong@asu.edu}
\email{rongchen@chla.usc.edu}
\affiliation{Department of Physics,
Arizona State University, Tempe, Arizona 85287, USA}
\affiliation{Laboratory of Applied Pharmacokinetics and Bioinformatics,
Children's Hospital Los Angeles, University of Southern California, Los Angeles, California 90027, USA}
\author{Kevin E. Schmidt}
\email{kevin.schmidt@asu.edu}
\affiliation{Department of Physics,
Arizona State University, Tempe, Arizona 85287, USA}
\date{\today}

\begin{abstract}
We describe a path-integral ground-state
quantum Monte Carlo method for light nuclei
in continuous space.
We show how to efficiently update and sample
the paths with spin-isospin dependent and spin-orbit interactions.
We apply the method to the triton and $\alpha$
particle using
both
local chiral interactions with next-to-next-to-leading-order %(N$^2$LO)
and the Argonne interactions. For operators, like the
total energy, that commute with the
Hamiltonian, our results agree with Green's function Monte Carlo
and auxiliary field diffusion Monte Carlo calculations. For operators
that do not commute with the Hamiltonian and for Euclidean response
functions,
the path-integral formulation allows straightforward calculation without
forward walking or the increased variance typical of diffusion methods.
We demonstrate this by calculating
density distributions, root-mean-square radii, and
Euclidean response functions for single-nucleon couplings.
\end{abstract}

\maketitle
%\tableofcontents % turn it off can significantly decrease compiling time.

\allowdisplaybreaks

\section{Introduction}\label{intro}

Real space path-integral quantum Monte Carlo
methods for calculating the properties of many-body systems
with spin independent interactions
\cite{Ceperley95a,Schmidt2000JCP,Baroni99a},
are often the methods of choice to calculate the ground-state
expectation values of operators such as the one- and two-body
density distributions, response functions, etc. that do not
commute with the Hamiltonian.

Real-space nuclear quantum Monte Carlo calculations performed
with diffusion Monte Carlo methods, such
as Green's
function Monte Carlo (GFMC) method \cite{Carlson87a} or the auxiliary
field diffusion Monte Carlo (AFDMC) \cite{Schmidt99a},
sample the product of the ground-state wave function
and the adjoint of a trial function.
These methods can accurately
calculate the ground-state energy and expectations of
other operators that commute with the Hamiltonian
using a mixed expectation value.

The calculation of other operators requires
additional
calculational steps, such as forward walking
\cite{Runge92PRB,Casulleras1995PRB,Samaras1999ForwardwalkingGF},
in order to calculate their
ground-state expectation. Forward walking methods typically have
higher variance which can limit the length of the forward walked
path and therefore the accuracy.

Many operators whose ground-state expectation is desired
do not commute with the Hamiltonian. Two examples are
the root-mean-square (RMS) radii and the particle number density. Additionally,
calculating
response functions where the system is propagated between the application
of
two operators require similar forward walking techniques
with diffusion
Monte Carlo methods.

Path-integral Monte Carlo (PIMC) methods \cite{Ceperley95a,Schmidt2000JCP},
on the other hand, sample and store the entire path. The energy calculation
is typically more time consuming, however, the advantage is that
the operators can be readily inserted anywhere
along the path.
The ground-state expectation arbitrary operators or response
functions is straightforwardly implemented.

In this paper we show how to implement
a real-space path-integral method for realistic nuclear
Hamiltonians. In order to give a proof of principle demonstration and
investigate path sampling methods,
we avoid the fermion sign or phase problem by looking at
s-wave nuclei ($A\leq 4$) only, where the sign or phase problem is
weak, and the calculations converge to their ground-state values
before any substantial sign or phase problem manifests itself.

We use the local chiral interaction
with
next-to-next-to-leading-order \cite{Gerzerlis14a,Lynn2017PRC,Lonardoni2018PRC}
(N$^2$LO),
and the Argonne $v'_6$ (AV6$'$) and $v'_8$ (AV8$'$) interactions \cite{WiringaPieper02a}.
We do not include three-body interactions, but they are
straightforward to include in future calculations.

We calculate the ground-state
energy, which can be directly compared with GFMC and AFDMC
results \cite{WiringaPieper02a,Lonardoni2018PRC}, and give
path-integral results for the RMS radius,
particle density distribution and several Euclidean response functions
for single-nucleon couplings \cite{Carlson94a} whose operators do not
commute with the Hamiltonian.

The paper is organized as follows.
In Sec. \ref{TheoFrame}, we describe
the framework for PIMC calculations.
We introduce the Hamiltonian,
the model space,
the trial wave function,
the short-time approximated propagator.
We describe
the error structure of our calculation
which will be used to extrapolate to zero time step.
In Sec. \ref{compalgorithm},
we describe the PIMC simulations and
how to update the path efficiently.
In Sec. \ref{secMCsampling},
we investigate various Metropolis Monte Carlo sampling moves and strategies
to sample the paths.
In Sec. \ref{RsDs} we
show results for the ground-state energy, RMS radii, single-particle
number density
and Euclidean response functions for single-nucleon couplings.
Section \ref{summary} gives
a summary and outlook for possible future calculations.

\section{Theoretical Framework}
\label{TheoFrame}

PIMC methods obtain the ground-state expectation value of a
Hermitian operator $\hat{O}$ using the matrix elements
\begin{equation}
\langle \hat{O} \rangle
= \frac{ \langle \Psi_T| e^{-H \tau_1}  \hat{O} e^{-H \tau_2 } | \Psi_T \rangle  }
{ \langle \Psi_T| e^{-H \tau } | \Psi_T \rangle}
\label{PIMCopexpdetail}
\end{equation}
where $\tau_1$ and $\tau_2$ are imaginary times and the
total imaginary time is $\tau=\tau_1+\tau_2$.
$\Psi_T$ is a trial wave function which serves as the initial
and final
states of the path.
$H = T+V$ is the Hamiltonian of the system where $T$ and $V$ are kinetic and potential energy operators.

For $\tau_1=\tau_2=0$,
the Eq.(\ref{PIMCopexpdetail}) is equivalent
to a variational Monte Carlo calculation with trial state
$|\Psi_T\rangle$.

For $\tau$ large enough, $e^{-H \tau} | \Psi_T \rangle$
projects out the ground state $| \Phi_0 \rangle$, i.e.,
$\lim\limits_{\tau \rightarrow \infty } e^{-H \tau} | \Psi_T  \rangle
\propto | \Phi_0 \rangle$.
So Eq.(\ref{PIMCopexpdetail}) becomes the ground-state expectation
value of $\hat{O}$,
\begin{equation}
\langle \hat{O} \rangle  = \frac{ \langle \Phi_0 | \hat{O}   | \Phi_0  \rangle }
{ \langle \Phi_0 | \Phi_0  \rangle  }
\label{PIMCopgndexp}
\end{equation}
when $\tau_1$ and $\tau_2$ are large.

Mixed estimates, which are the exact ground-state expectations for
the Hamiltonian and
operators commute with the Hamiltonian, can be calculated by
taking
one of the
$\tau_1$ or $\tau_2$ large and the other zero.

The propagator $e^{-H \tau_i}$ is sampled by writing
\begin{equation}
e^{-H \tau_i}=  \left ( e^{-H \Delta \tau }\right )^{N_i}
\,. \label{wholepropagator}
\end{equation}
and using a Trotter breakup of $e^{-H\Delta \tau}$
to approximate the short-time propagator.

In order to calculate the ground-state properties, we must
use a sufficiently large $\tau_i$ such that $e^{-H \tau_i}$ can project out the ground state $|\Phi_0\rangle$. The time step $\Delta \tau$
is chosen
small enough such that Trotter breakup errors can be either ignored or
extrapolated out. A typical calculation is formulated as
\begin{equation}
\langle \hat{O} \rangle
= \frac{\langle \Psi_T| e^{-H \Delta \tau }...e^{-H \Delta \tau } \hat{O} e^{-H \Delta \tau } ...e^{-H \Delta \tau }| \Psi_T \rangle  }
{\langle \Psi_T| e^{-H \Delta \tau }...e^{-H \Delta \tau } e^{-H \Delta \tau } ...e^{-H \Delta \tau } | \Psi_T \rangle} .
\label{PIMCopexpdetailshorttime}
\end{equation}

Since $T$ and $V$ do not commute,
the Trotter breakup for the short-time propagator $U(\Delta \tau )$
has a time-step error
\begin{equation}
U(\Delta \tau ) = e^{-H \Delta \tau } + \mathcal{O}(\Delta \tau^k),
\end{equation}
of order $\Delta \tau^k$.
Different choices of $U(\Delta \tau )$ give different order for $k$.

Our calculations give
$\langle \hat{O}(\Delta \tau ) \rangle$, an approximation of the true expectation value $\langle \hat{O} \rangle$ in Eq.(\ref{PIMCopexpdetailshorttime}),
\begin{equation}
\langle \hat{O} (\Delta \tau ) \rangle
= \frac{\langle \Psi_T| [U(\Delta \tau )]^{N_1} \hat{O} [U(\Delta \tau )]^{N_2}| \Psi_T \rangle  }
{\langle \Psi_T| [U(\Delta \tau )]^{N_1+N_2} | \Psi_T \rangle} .
\label{PIMCopexpdetailshorttimeapprox}
\end{equation}
In the limit where the time step $\Delta \tau$ becomes zero,
\begin{equation}
\lim_{\Delta \tau      \rightarrow 0}\langle \hat{O} (\Delta \tau ) \rangle
= \langle \hat{O} \rangle .
\end{equation}
To estimate the errors,
we calculate $\langle \hat{O}(\Delta \tau ) \rangle$ for different time step $\Delta \tau$ and extrapolate to $\Delta \tau=0$ to find the true expectation value $\langle \hat{O} \rangle$.
In Eqs.(\ref{PIMCopexpdetailshorttime})
and
(\ref{PIMCopexpdetailshorttimeapprox}),
for large-enough $N_1$ and $N_2$, the numerator and denominator are real,
and in the Monte Carlo results, we keep just the real part.

An advantage of PIMC is that since the entire path is sampled,
it can directly deal with operators which do not commute with $H$,
while diffusion-based methods like GFMC and AFDMC cannot do so without
additional techniques such as forward walking \cite{Runge92PRB}.
As we
can see from Eq.(\ref{PIMCopexpdetailshorttime}),
a PIMC calculation
depends on not only the operator $\hat{O}$,
but also the Hamiltonian ${H}$, the trial wave function $\Psi_T$,
and the short
time propagator $e^{-H \Delta \tau }$. In the next sections we will
describe each of these.

\subsection{Hamiltonian}
\label{secHamilton}

We choose
the Hamiltonian $H$ for our calculations to include just two-body
potentials
\begin{equation}
H = \sum_{i=1}^A \frac{\bm{p}_i^2}{2m}  + \sum_{i<j}V_{ij}, \label{PIMCHamilton}
\end{equation}
where $A$ is total number of nucleons in the system,
in the position representation
$\bm{p}_i=-i\hbar \nabla_i$,
and $\frac{\hbar^2}{2m}$ is chosen as $20.375$ \textrm{fm}$^2$.
Since our PIMC calculation is a proof of principle benchmark test,
we did not include three-body interaction in this work.
The complete two-body interaction for a given $ij$ pair of particles, $V_{ij}$, is composed of the nucleon-nucleon strong
interaction $V^\textrm{NN}_{ij}$ and the electromagnetic force $V^\textrm{EM}_{ij}$,
\begin{equation}
V_{ij}=V_{ij}^\textrm{NN}+V^\textrm{EM}_{ij}.
\label{Vij}
\end{equation}

We use the local chiral interaction with N$^2$LO \cite{Gerzerlis14a,Lynn2017PRC,Lonardoni2018PRC}, the Argonne $v'_6$ (AV6$'$)
and $v'_8$ (AV8$'$) interactions \cite{WiringaPieper02a}.
The nucleon-nucleon interaction can be written as,
\begin{equation}
V^\textrm{NN}_{ij}=\sum_{p=1}^8 v_p(r_{ij}) O_{ij}^p , \label{AV6'}
\end{equation}
where $r_{ij}$ is the length of $\bm{r}_{ij}$, $v_p(r_{ij})$ is the radial function for the $p^{\textrm{th}}$ operator, with different functions
for the chiral and Argonne interactions.
The first six of the operators $O_{ij}^p$ are the same for all the
potentials we are using.
They are 1,
$\bm{\tau }_i \cdot \bm{\tau }_j$,
$\bm{\sigma }_i \cdot \bm{\sigma }_j$,
$\bm{\sigma }_i \cdot \bm{\sigma }_j \bm{\tau }_i \cdot \bm{\tau }_j$,
$S_{ij}$
and $S_{ij} \bm{\tau }_i \cdot \bm{\tau }_j$,
where $\bm{\sigma }$ and $\bm{\tau }$ are spin and isospin operators
and $S_{ij}$ is the tensor force.
The local chiral N$^2$LO interaction has the additional
seventh operator which is a spin-orbit term,
\begin{equation}
O_{ij}^7 = \bm{L}      \cdot \bm{S},
\label{vlsrij}
\end{equation}
where $\bm{L}$ is the relative angular momentum and $\bm{S}$ is the total spin \cite{Lonardoni2018PRC}.
For the AV8' interaction, besides the first seven operators,
it has the eighth operator which is a spin-orbit term coupled with the isospin term,
\begin{equation}
O_{ij}^8 = \bm{L} \cdot \bm{S} {\bm\tau}_i\cdot{\bm\tau}_j ,
\label{vlsttrij}
\end{equation}

For the electromagnetic force $V^\textrm{EM}_{ij}$, here we just consider the Coulomb force term $v^\textrm{C}(r_{ij})$ between proton pairs.

\subsection{Model space}
\label{secModSpace}

The interaction used in this paper will not change $Z$,
the total number
of protons, so the number of possible isospin states
is $\frac{A!}{Z!(A-Z)!}$.
The tensor part $S_{ij}$ can flip spins, so all $2^A$ spin states
are allowed, giving
$ N_{\rm{tot}}= \frac{A!}{Z!(A-Z)!} 2^A$
for the total number of spin-isospin basis states
\footnote{
In our PIMC code, the spin, isospin states are written using
a binary representation \cite{Koonin91},
and we label the $A$ particles from 0 to $A-1$. The various Pauli
matrix operators for the spin and isospin are implemented by bit flips
and exchanges along with multiplication of the corresponding coefficients.
}.

$^4$He then
has $N_{\rm{tot}}=6      \times      16=96$ spin-isospin basis states.
We write
these basis states as $| S \rangle$ and $S$ takes 96 values,
\begin{equation}
| S \rangle      \equiv       | s_1\rangle     | s_2 \rangle   | s_3 \rangle      | s_4 \rangle      \equiv      | s_1 s_2 s_3 s_4 \rangle, \label{basisS}
\end{equation}
where $s_i$ means the spin-isospin state of particle $i$,
it can be any state from neutron spin up $|n \uparrow \rangle$,
neutron spin down $|n \downarrow\rangle$,
proton spin up $|p \uparrow\rangle$,
and proton spin down $|p \downarrow\rangle$.
We denote the spatial configuration of a system with $A$ particles as $R$
such that $ R \equiv ( \bm{r}_1, \bm{r}_2, \ldots, \bm{r}_A  )$,
where $\bm{r}_i=(x_i,y_i,z_i)$ is the coordinates of particle $i$.
The spatial configuration of $^4$He can also be written as a state $|R \rangle$,
\begin{equation}
| R \rangle      \equiv       | \bm{r}_1\rangle    | \bm{r}_2 \rangle
 | \bm{r}_3 \rangle     | \bm{r}_4 \rangle
\equiv | \bm{r}_1 \bm{r}_2 \bm{r}_3 \bm{r}_4 \rangle. \label{basisR}
\end{equation}

The basis states $| RS \rangle$ are
\begin{equation}
| RS \rangle = | R \rangle   | S \rangle
\equiv      | \bm{r}_1 \bm{r}_2 \bm{r}_3 \bm{r}_4 \rangle
| s_1 s_2 s_3 s_4 \rangle , \label{basisRS}
\end{equation}
with the
corresponding identity operators
$ \int dR | R \rangle  \langle R | =  1$,
$\sum_S | S \rangle  \langle S | =  1$, and
$\sum_S \! \! \int \!\! dR | RS \rangle  \langle RS | = 1$.

If we ignore
the Coulomb interaction, then the potentials conserve total isospin.
Often GFMC calculations include the isospin breaking terms as a perturbation
and work in a good total isospin basis,
along with exploiting time-reversal
invariance for the integer total spin case.
This reduces
the calculational basis size.
We chose instead to include the isospin
breaking interactions in the propagator and did not exploit
time-reversal symmetry.

\subsection{Wave function}
\label{secWavefunc}

We use
a trial state $| \Psi_T \rangle$ of the form \cite{CarlsonRMP15a},
\begin{equation}
|\Psi_T \rangle = \mathcal{F} |\Phi \rangle
= {\mathcal S} \prod _{i<j} F_{ij} |\Phi \rangle , \label{PsiTstateF}
\end{equation}
where $| \Phi \rangle$ is the model state,
$\mathcal{F}$ is the correlation operator which is a product
of two-body correlation operators $F_{ij}$,
\begin{equation}
F_{ij} =  \sum_{p=1}^6 f_{ij}^p O_{ij}^p \, , \label{coropFij}
\end{equation}
where $O_{ij}^p$ are the AV6' operators and $f_{ij}^p $
is the corresponding correlation function and
$\mathcal{S}$ is a symmetrization operator that acts only on the
correlations to guarantee a properly antisymmetric state,

For $A\leq 4$, an s-wave model state can be constructed from a constant
spatial function multiplying antisymmetric combinations of spin-isospin
states and all spatial dependence is included
as in Ref. \cite{Lomnitz1981NPA}.
We calculate the correlations
by solving the two-body differential equations as described in Ref.
\cite{CarlsonRMP15a} and Ref.
\cite{Lomnitz1981NPA} and adjust the parameters using the variational method.
$|\Phi\rangle$ is chosen to have the correct quantum numbers for the desired
state.

For $A\leq 4$,
$| \Phi \rangle$ can be decomposed by a spatial part $|\Phi_\textrm{R} \rangle$ and a spin-isospin part $| \Phi_\textrm{S} \rangle$ such that
\begin{equation}
|\Phi \rangle =  |\Phi_\textrm{R} \rangle  | \Phi_\textrm{S} \rangle,
\label{Phi}
\end{equation}
where
$|\Phi_\textrm{R} \rangle$ is symmetrized which can be chosen as
$ |\Phi_\textrm{R} \rangle = \int dR  | R \rangle$
such that
$\langle R'|\Phi_\textrm{R} \rangle = \int dR \delta (R'-R)=1$.

For example, for $^4$He, the spin-isospin model state is simply
\begin{equation}
| \Phi_\textrm{S} \rangle
= \mathcal{A}   | n \uparrow n \downarrow p \uparrow p \downarrow  \rangle
= \sum_{N=1}^{96} \phi_N | N \rangle . \label{PhiS}
\end{equation}
$\mathcal{A}$ is the antisymmetrization operator, and $\phi_N$
is either $-1$ or $1$ depending on the antisymmetrization for each of
the 24  basis states, and 0 for the rest 72 basis states.
The wave function $\langle RS | \Phi \rangle$ becomes
\begin{equation}
\langle RS | \Phi \rangle
=\sum_{N=1}^{96} \phi_N \langle S | N \rangle
\,.
\label{RSPhi}
\end{equation}
The trial wave function is
\begin{equation}
\langle R S | \Psi_T  \rangle \! = \!
\left\langle S\left |
{\mathcal S}  \prod     _{i<j}
\left[
\sum_{p=1}^6   f_{ij}^p (r_{ij})  O_{ij}^p
\right]
\right |\Phi_\textrm{S}\right \rangle
\label{RSPsiT}
\end{equation}
where we sample the order of the correlations to apply the symmetrization
operator ${\mathcal S}$ as in Refs.
\cite{Lomnitz1981NPA}
and
\cite{CarlsonRMP15a}.
We use a superscripts $l$ and $r$
on the states and products
to denote a particular sampled order for the left and right
trial functions of our path integral,
$\langle \Psi^l_T |$ from $\langle \Psi_T |$
and $| \Psi^r_T \rangle$ from $| \Psi_T \rangle$ with
\begin{eqnarray}
\langle \Psi_T | RS \rangle &=& \sum  _l\langle \Psi^l_T |  RS \rangle , \label{lorderdef}\\
\langle RS | \Psi_T \rangle &=& \sum_r \langle RS | \Psi^r_T \rangle, \label{rorderdef}
\end{eqnarray}
and
\begin{eqnarray}
\! \! \! \! \! \! \langle R S | \Psi_T^r  \rangle &=& \!
\left\langle S \left|
\sideset{}{^r}
 \prod     _{i<j}
\left[
\sum_{p=1}^6   f_{ij}^p (r_{ij})  O_{ij}^p
\right]
\right | \Phi_\textrm{S}\right \rangle
, \label{RSPsiTr} \\
\! \! \! \! \! \!  \langle \Psi_T^l | R S \rangle &=& \!
\left \langle \Phi_\textrm{S}\left |
\sideset{}{^l} \prod     _{i<j}
\left[
\sum_{p=1}^6  {f_{ij}^p}^* (r_{ij})   O_{ij}^p
\right]
\right| S \right\rangle .
\label{RSPsiTl}
\end{eqnarray}

\subsection{Propagator}
\label{propagator}
We write
the free-particle propagator as
\begin{equation}
%\begin{split}
G^f_{R'R}       =
\langle R' | e^{-T \Delta \tau }  | R \rangle
=
\left( \frac{m}{2\pi \Delta \tau \hbar^2} \right)^{\frac{3A}{2}}
 e^{  -\frac{ (R'-R)^2 }{2 \Delta \tau \frac{\hbar^2}{m}}  } ,
\label{propagatorFree}
%\end{split}
\end{equation}
where $ (R'\!-\!R)^2 = \sum\limits_{i=1}^A  (\bm{r}_i'-\bm{r}_i)^2$,
and the potential part of the propagator, without the spin-orbit interaction,
as
\begin{align}
\hspace{-0.5em} U_V(R,\tfrac{\Delta t}{2}) \!= \!\!
\ e^{\frac{-\mathcal{V}_{A-1 A} \Delta t}{2}}
e^{\frac{-\mathcal{V}_{A-2 A} \Delta t}{2}}
...
e^{\frac{-\mathcal{V}_{13}  \Delta t}{2}}
e^{\frac{-\mathcal{V}_{12} \Delta t}{2}}
\label{propagatorv6half}
\end{align}
where $\mathcal{V}_{ij}$ contains the pair potential with the first six operators
in Eq.(\ref{AV6'}) plus the electromagnetic force,
\begin{equation}
\mathcal{V}_{ij}  =  \sum_{p=1}^6 v_p(r_{ij}) O_{ij}^p + V^\textrm{EM}_{ij}(r_{ij}) .
\label{PER}
\end{equation}
We choose a fixed order for the pair-potential exponentials.
The operator $U_V^\dagger(R,\tfrac{\Delta t}{2})$ reverses the order
of these exponentials. In the calculations,
each of these exponentials is rewritten as
${e}^{ -\sum\limits_{p=1}^6 v_p(r_{ij}) O_{ij}^p \Delta \tau}  =
 \sum\limits_{p=1}^6 u^p_{ij} (r_{ij}) O^p_{ij}$,
where we can solve for the coefficients $u^p_{ij} (r_{ij}) $
given the $r_{ij}$.

Without a spin-orbit interaction, the short-time propagator is
\begin{equation}
\label{propagatorNoV6}
\begin{split}
& \langle R' S' |U(\Delta \tau)|R S\rangle\\
&=
\left \langle S'\left |
U_V^\dagger(R',\tfrac{\Delta t}{2})
U_V(R,\tfrac{\Delta t}{2})
\right | S \right \rangle
G^f_{R'R}
 \,.
\end{split}
\end{equation}

Since the spin-orbit interaction is relatively weak,
we include it at linear order in the time step. This avoids
needing to include counter terms \cite{CarlsonRMP15a}.
Operating
the $\bm{p}_j$ operators on the free-particle propagator replaces them
with  $i m \frac{({\bm r}_j'-{\bm r}_j)}{\hbar \Delta \tau}$.
The short-time propagator with spin-orbit interactions becomes
\begin{align}
\label{propagatorN2LORS}
\begin{split}
&\langle R'S' |U(\Delta \tau)|RS\rangle
\\
&=
\langle S'|
U_V^\dagger(R',\tfrac{\Delta \tau}{2})
\mathcal{G}(R',R)
U_V(R,\tfrac{\Delta \tau}{2})
|S \rangle
G^f_{R'R}\,,
\end{split}
\end{align}
with
\begin{align}
\label{defdownarroweff}
\begin{split}
\mathcal{G}(R',R) &=
1 +
 \frac{m}{4i  \hbar^2}
\sum_{i<j} \left [v_7(r'_{ij})+v_8(r'_{ij}){\bm\tau}_i\cdot{\bm\tau}_j\right]
\cdot
\\
&
\left[ \bm{r}'_{ij} \times
\Delta \bm{r}^{R',R}_{ij} \cdot (\bm{\sigma }_i+\bm{\sigma }_j) \right] ,
\end{split}
\end{align}
where $r'_{ij}$ is the distance between particle $i$ and $j$ in
configuration $R'$,
$\bm{r}'_{ij} \equiv \bm{r}_i-\bm{r}_j$ in $R'$,
$\Delta \bm{r}^{R',R}_{ij}\equiv \Delta \bm{r}^{R',R}_i-\Delta \bm{r}^{R',R}_j$
where $\Delta \bm{r}^{R',R}_{i(j)}      \equiv      \bm{r}^{R'}_{i(j)} - \bm{r}^R_{i(j)} $,
and the symbol $\bm{r}^{R}_{i(j)}$
means the $\bm{r}_{i(j)}$ in configuration $R$.

\subsection{Error estimation}
\label{secErrEst}

We use the $U(\Delta \tau )$ in Eq.(\ref{propagatorN2LORS})
as the short-time propagator for local chiral N$^2$LO interaction
and the AV8' interaction.
It gives
a time-step error for the path which is linear in the time step.
We fit the coefficient $C_1$ to this error for short times,
\begin{equation}
\langle \hat{O} (\Delta \tau) \rangle
= \langle \hat{O} \rangle + C_{1} (\Delta \tau),
\label{ON2LOextrapolation}
\end{equation}
to extrapolate to the zero time-step limit.

For AV6$'$ we use $U(\Delta \tau )$ in Eq.(\ref{propagatorNoV6})
and we can immediately find in this case $U(\Delta \tau ) U(-\Delta \tau )=1$.
The difference between $e^{-H \Delta \tau }$ and $U(\Delta \tau )$ only
contains odd order terms in $\Delta \tau$ \cite{SchmidtLee95a,Suzuki1991PLA}.
The time-step error in the full path is extrapolated to zero by
fitting to
\begin{equation}
\langle \hat{O} (\Delta \tau) \rangle
= \langle \hat{O} \rangle
+  C_{2} (\Delta \tau^{2})\,.
\label{Oextrapolation}
\end{equation}

\section{Path-integral form}
\label{compalgorithm}
%\subsection{Sums and integrals over probabilities}
%\label{Recast}
We write Eq.(\ref{PIMCopexpdetailshorttimeapprox}) in a form suitable
for Monte Carlo calculations.
The total time $\tau=\tau_1+\tau_2=N \Delta \tau$, with $\Delta \tau$
the time step.
%We can insert the identity operators described in Sec. \ref{secModSpace} between all the $U(\Delta \tau )$ in Eq.(\ref{PIMCopexpdetailshorttimeapprox}).
%Then by using Eq.(\ref{propagatorN2LORS}) for each $\langle R_I S_I |  U(\Delta \tau )   | R_J S_J \rangle$, and then remove all the identity operator such as $ \sum_{S_I} | S_I \rangle \langle S_I | = 1$, we are able to obtain,
The path integral becomes
\begin{widetext}
\begin{flalign}
\langle \hat{O} (\Delta \tau )\rangle &=
\frac{ \! Re \displaystyle  \sum_{S_0,S_N} \!
 \int \! \mathcal{D} \mathcal{R}   \langle \Psi_T| R_0 S_0\rangle
\langle R_0 S_0 | [U(\Delta \tau )]^{N_1} \hat{O} [U(\Delta \tau )]^{N_2}| R_{N} S_{N} \rangle
\langle R_{N} S_{N}  | \Psi_T \rangle  }
{ Re \! \displaystyle \sum_{S_0,S_N} \!
 \int \! \mathcal{D} \mathcal{R}  \langle \Psi_T| R_0 S_0\rangle
\langle R_0 S_0 | [U(\Delta \tau )]^{N} | R_N S_N \rangle
\langle R_{N} S_{N}  | \Psi_T \rangle } ,
\label{OABP1}
\end{flalign}
\end{widetext}
where
we assume that the times are
long enough to make the numerator and denominator real, as noted above,
and take the real parts in the Monte Carlo calculations.
The symbol $\mathcal{R}$ denotes all of
the spatial configurations $\{R_0, R_1, \ldots, R_N \}$,
and we call each of the $R_I$ a bead.
The integral of $\mathcal{R}$ is the spatial integral over all the configurations $R_I$,
i.e., $ \int \mathcal{D} \mathcal{R}      \equiv       \prod_{I=0}^N \int d R_I $.

In addition to the position integrals,
since we sample the order of the operators
in the trial functions, we write
the sampled left and right orders as $l$ and $r$, as in Eqs. (\ref{lorderdef}) and (\ref{rorderdef}).
Eq. (\ref{OABP1}) is then in the form
\begin{eqnarray}
\langle \hat{O} (\Delta \tau ) \rangle
&=&
\frac{\displaystyle \sum_l \sum_r \int \mathcal{D} \mathcal{R} A_{lr} (\mathcal{R}) P_{lr} (\mathcal{R}) }
{\displaystyle \sum_l\sum_r \int \mathcal{D} \mathcal{R} B_{lr} (\mathcal{R}) P_{lr} (\mathcal{R})  }
\nonumber\\
&=&
\frac{\langle A_{lr}(\mathcal{R}) \rangle }{\langle B_{lr}(\mathcal{R}) \rangle}
\bigg|_{ \{l,r,\mathcal{R}\}      \in P_{lr}(\mathcal{R})  } .
\label{OABP}
\end{eqnarray}

In Eq.(\ref{OABP}),
the $A_{lr} (\mathcal{R})$ and $B_{lr} (\mathcal{R})$
are real functions which can be written as
$A_{lr} (\mathcal{R}) = \frac{Re[ g^V_{lr,M}(\mathcal{R})]}{|Re[ f^V_{lr}(\mathcal{R})]|  }$
and
$B_{lr} (\mathcal{R}) = \frac{Re[ f^V_{lr}(\mathcal{R})]}{|Re[ f^V_{lr}(\mathcal{R})]|  }$,
with $B_{lr} (\mathcal{R})$ either 1 or $-$1 indicating the weak sign
problem for the $A \leq 4$ nuclei.
$P_{lr} (\mathcal{R})$ is the normalized probability distribution,
\begin{equation}
P_{lr} (\mathcal{R}) =
\frac{|Re[ f^V_{lr}(\mathcal{R}) g^F(\mathcal{R})   ]|}{ \mathcal{N}  },  \label{PlrR_detail}
\end{equation}
with
$ \sum_{l}\sum_{r} \int d \mathcal{R} P_{lr}(\mathcal{R}) =1$.
As usual, the normalization factor
$ \mathcal{N}$ cancels in the Metropolis algorithm implementation.

The detailed forms and calculations of the functions
$g^F(\mathcal{R})$,
$g^V_{lr,M}(\mathcal{R})$,
and $f^V_{lr}(\mathcal{R})$ are presented in
Eqs. (\ref{gF}), (\ref{gVRlrdef2}), and (\ref{fVRlrN2LOdef2}) in
Appendix. \ref{secCalcPath} which describes how the path is calculated.
The calculations of the path updating is presented in Appendix. \ref{secUpdatePathStrategy}.
More details can be found in Chapter 2 in Ref. \cite{myphdthesis}.

With Eq.(\ref{PIMCopexpdetailshorttimeapprox}) written
in the form of Eq. (\ref{OABP}), it can now be calculated
by sampling the probability distribution
$P_{lr}(\mathcal{R})$, and averaging the numerator and denominator
$ A_{lr} (\mathcal{R}) $ and $ B_{lr}(\mathcal{R})$ in Eq.(\ref{OABP}),
along with the statistical errors.

\section{Monte Carlo Sampling}
\label{secMCsampling}

Our PIMC is based on the standard
Metropolis method \cite{Ceperley95a,Metropolis53a,HammondMCbook}.
We calculate $\langle F \rangle$, the expectation value of the function $F(s)$,
as
\begin{equation}
\langle F \rangle = \sum_s F(s) \pi (s),
\label{Fave}
\end{equation}
where $\pi (s)$ is a normalized probability distribution such that $\sum_s \pi (s)=1$;
it describes the probability for state $s$ to occur, where $s$ represents
the set of sampled variables. In the standard Metropolis method
we propose a transition from the state $s$ to a new state $s'$
with probability $T(s     \rightarrow s')$. We accept the new state
with the Metropolis probability
\begin{equation}
A(s     \rightarrow s')=
\min \left[  1, \frac{\pi (s')T(s'     \rightarrow s)}{\pi (s) T(s     \rightarrow s')} \right] ,
\label{Accptequation}
\end{equation}
satisfying detailed balance.
Below we will describe several transition probabilities that we have used
to implement an efficient path integral sampling for the nuclear problem.

\subsection{Metropolis method in PIMC}
\label{secMetroPIMC}
We take $\pi(s)$ in Eq.(\ref{Fave})
to be $P_{lr} (\mathcal{R})$ in Eq.(\ref{PlrR_detail}).
The state $s$ is then a particular choice of the left and right
correlation operator orders in the trial function and
the bead positions that describe the path
\{$l,r,\mathcal{R}$\}.

We separately sample the left and right trial wave-function correlation
operator order and the bead positions.

For the sampling of the operator orders in the trial wave function, we write
\begin{equation}
T(s     \rightarrow s') = T_{lr \rightarrow l'r'}.
\end{equation}
We randomly choose new permutations of the orders,
so $T_{lr \rightarrow l'r'}$ and $T_{l'r' \rightarrow lr}$ are equal
and the acceptance probability is

\begin{equation}
A (s     \rightarrow s') \! = \!
\min \! \left[  1,
\frac{ f^V_{l'r'}(\mathcal{R}) g^F(\mathcal{R})  }
{ f^V_{lr}(\mathcal{R}) g^F(\mathcal{R})  }
\right] .
\end{equation}
Since only the ordering of the trial wave-function correlations changes,
these calculations are independent of the length of the path. Since
the commutators of the correlations are typically small, these
moves are usually accepted.

We propose new beads positions $\mathcal{R}'$
with the transition probability
\begin{equation}
T(s     \rightarrow s')=
T_{lr}(\mathcal{R} \rightarrow \mathcal{R}'),
\end{equation}
where $T_{lr}(\mathcal{R} \rightarrow \mathcal{R}')$ samples the new positions $\mathcal{R}'$ for the path beads, given the new accepted order $lr$.
We use several different methods described below to efficiently sample
the paths.

\subsection{Multilevel sampling}
\label{secGaussianSample}

We write
the proposed transition probability $T_{lr}(\mathcal{R}      \rightarrow \mathcal{R}')$ and $T_{lr}(\mathcal{R}'      \rightarrow \mathcal{R})$ in Sec \ref{secMetroPIMC} in a way such that,
\begin{equation}
g^F(\mathcal{R}') T_{lr}(\mathcal{R}'      \rightarrow \mathcal{R}) = g^F(\mathcal{R}) T_{lr}(\mathcal{R}      \rightarrow \mathcal{R}')   ,
\label{gtnewgtold}
\end{equation}
and the corresponding acceptance rate becomes,
\begin{equation}
A(s     \rightarrow s')=
\min \left[  1, \frac{ f^V_{lr}(\mathcal{R}') }{ f^V_{lr}(\mathcal{R})  }\right] .
\label{Accptequationsimp}
\end{equation}

We use the Gaussian propagators in $g^F(\mathcal{R}')$ to sample such a new proposed free-particle path $\mathcal{R}'$.
There are many ways to do this. Here, since we wished to
test the implementation of the bisection method \cite{Ceperley95a} commonly
used for central hard-core interactions, we sample the points using
what we call multilevel sampling.

From the total $N+1$ beads on the imaginary time from bead $0$ to bead $N$,
we randomly pick $2^n+1$ consecutive beads from bead $I$ to bead $I+2^n$,
where $n$ is a positive integer.
Here we refer to these beads
from bead $I$ to bead $I+2^n$ as $R_0$ to $R_{2^n}$, with
configurations represented by
\begin{equation}
\mathcal{R}=\{R_0, R_1, R_2, \ldots, R_{2^n-1}, R_{2^n} \}. \label{mathcalRmulbeads}
\end{equation}

We keep $R_0$ and $R_{2^n}$ unchanged
and propose the new configurations for the beads in between them.
The relevant new proposed beads at each level are,
\begin{eqnarray}
\textrm{Level } 0 &:&  R_0, R_{2^n} .     \nonumber \\
\textrm{Level } 1 &:&  R'_{\frac{1}{2}2^n}  .     \nonumber \\
\textrm{Level } 2 &:& R'_{\frac{1}{4} 2^n} , R'_{\frac{3}{4} 2^n} .     \nonumber \\
 &\vdots &        \nonumber \\
\textrm{Level } k &:&  R'_{\frac{1}{2^k} 2^n} , R'_{\frac{3}{2^k} 2^n}, \ldots, R'_{\frac{2^k-1}{2^k} 2^n} .   \nonumber \\
 &\vdots &        \nonumber \\
\textrm{Level } n &:& R'_{1}, R'_{3}, \ldots, R'_{2^n-1} .        \nonumber
\end{eqnarray}

We denote bead $M$ as the bead located in the middle of bead $L$ and bead $N$.
We propose the new beads $R'_M$ at each level according to
the average position of $R'_L$ and $R'_N$ from the previous level,
plus a random Gaussian number vector $\bm{g}$
whose standard deviation at each dimension is
$\sigma=\sqrt{ \frac{|N-L|}{2}\Delta \tau \frac{\hbar^2}{2m}  }$
and average is zero,
\begin{align}
 R'_M = \frac{ R'_{L} + R'_{N} }{2} + \bm{g}.
\end{align}
In this way, we are actually setting the proposed transition probability
$T_{lr}(\mathcal{R}  \rightarrow \mathcal{R}')$ as,
\begin{equation}
T_{lr}(\mathcal{R} \rightarrow \mathcal{R}')
= g^F(\mathcal{R}').
 \label{Tlrnewmbeads}
\end{equation}
This satisfies Eq.(\ref{gtnewgtold}) and we can use
Eq.(\ref{Accptequationsimp}) for the acceptance probability.

To calculate ${ f^V_{lr}(\mathcal{R}')}/{ f^V_{lr}(\mathcal{R})}$,
we define states
$| \psi^r_M (\mathcal{R}) \rangle $ and $| \psi^l_M (\mathcal{R})\rangle$ as shown in Eqs. (\ref{phirMRrecur}) and (\ref{philMRrecur})
in Appendix \ref{secUpdatePathStrategy},
as well as the corresponding probability amplitude.
$\phi^r_{M}(\mathcal{R},S)$ and $\phi^l_{M}(\mathcal{R},S)$,
as shown in Eqs. (\ref{phirMexpansion}) and (\ref{philMexpansion})
in Appendix \ref{secCalcPath}.
When we propose new moves from bead $I+1$ to bead $I+2^n-1$,
we calculate from $\phi^r_{I+2^n-1}(\mathcal{R}',S)$
to $\phi^r_{I+1}(\mathcal{R}',S)$,
using the recursion relations Eq. (\ref{phirMRrecur}) starting from
our previously calculated
$\phi^r_{I+2^n}(\mathcal{R},S)$.
From this we use Eq. (\ref{fVRlrN2LOdef2}) to calculate
$f^V_{lr}(\mathcal{R}')= \left\langle \psi^{l}_I (\mathcal{R}')
\left|
\mathcal{G}(\mathcal{R}'_{I},\mathcal{R}'_{I+1})
\right| \psi^{r}_{I+1} (\mathcal{R}') \right\rangle $,
and with the previously calculated
$ f^V_{lr}(\mathcal{R})$ find
${ f^V_{lr}(\mathcal{R}')}/{ f^V_{lr}(\mathcal{R})}$.
If the proposed move is accepted, then we update the rest of the beads.
From right to left we update from $\phi^r_{I}(\mathcal{R}',S)$ until
$\phi^r_{0}(\mathcal{R}',S)$, and from left to right we update from
$\phi^r_{I+1}(\mathcal{R}',S)$ until $\phi^r_{N}(\mathcal{R}',S)$.

We choose the number of beads to optimize the time for these moves. Since
updating is relatively expensive, we choose the number of beads to obtain
a relatively small acceptance rate as discussed later.

\subsection{Bead sampling}
\label{secregularsampling}

In order to make the total PIMC algorithm more robust,
we can occasionally add a variety of other moves in addition
to multilevel sampling alone
in order to ensure an independent sampling of the path.

We randomly select a set of beads. In these beads we move the constituent
particles with a displacement, either uniformly in a cube or
with a Gaussian distribution, around the their current
position. As expected either sampling method gives a comparable acceptance.

We also tried moves where we translated all of the particles
in all of the beads by the same $3A$-dimensional random vector
$\Lambda$, i.e.,
\begin{equation}
 R'_I=R_I+\Lambda,
 \label{shiftmove}
\end{equation}
which we call a shift.

\subsection{Sampling strategies}
\label{secSampStrat}

Besides the sampling methods described this section, we also tested using
the bisection method \cite{Ceperley95a} and reptation Monte Carlo method
\cite{Baroni99a,Baroni10a} generalized to spin/isospin dependence.

The proposed trial moves at each level in multilevel sampling are the
same as bisection method \cite{Ceperley95a}.
The advantage of the bisection method is that moves where the middle bead of the path is sampled in region where the potential is highly repulsive,
and therefore likely to be rejected, are rejected early to minimize
computations. We tried bisection but found that with these
softer nuclear potentials with spin/isospin dependence, the approximate
path values when the first intermediate beads was sampled, was not
a particularly good predictor of the final path's value. Early rejection
of the path did not give a more efficient method. Therefore our acceptance
probability is based only on the
entire path and there is only a single accept-reject step after the
entire path is constructed.

The reptation method samples the spatial configuration of the path by
adding some beads at one end of the path and removing some beads to
the other end of the path.  However, since we have spin and isospin
sums in the path calculation, removing and adding beads, even if just
one bead is involved, requires the recalculation of the whole path.
Therefore, after initial testing, we did not
pursue reptation moves; they are not efficient here.

For multilevel sampling, for the total $N+1$ beads along the path, we
pick two lengths of sets of beads: $n_1 \approx N/3$ and $n_2 \approx
N/6$.
About $80\%$ of the moves are proposed for $n_1$ beads and the
others are for $n_2$ beads.
The acceptance rate for $n_1$ beads sampling
is about $20\%$ and for $n_2$ beads is about $40\%$.  The $n_2$ beads case
are mainly used for sampling the beads including the 0th and the $N$th beads.

The bead sampling methods discussed above are included
but are not performed as often since they are not as efficient as the multilevel sampling.

About $90\%$ of the moves use multilevel sampling and the other $10\%$ are proposed by different kinds of bead sampling,
which include moving beads one by one, moving all the beads at the same time, and a shift of all the beads at the same time, as described in Sec. \ref{secregularsampling}.

\section{Results and Discussions}
\label{RsDs}

We have calculated
the ground-state energy, density distribution, root-mean-square radii, and Euclidean response functions for single-nucleon couplings.
The detailed expressions such as the $g^V_{lr,M}(\mathcal{R})$  for each of these quantities can be found in Ref. \cite{myphdthesis}.

\subsection{Ground-state energy}
\label{secGS}

The calculation of the ground-state energy $E_0$ can be written in
the form of Eq.(\ref{OABP}) as discussed in Sec. \ref{compalgorithm},
with the operator $\hat{O}$ being the Hamiltonian.
As indicated by
Eq. (\ref{gRlrOH}), we calculate the ground-state energy $E_0$ using
\begin{equation}
E_0
= \frac{
\frac{1}{2}  Re \! \left( \!   \langle \Psi_T| H e^{-H \tau }  | \Psi_T \rangle
\! + \!
 \langle \Psi_T|  e^{-H \tau } H | \Psi_T \rangle
\! \right)
 }
{  Re \langle \Psi_T| e^{-H \tau } | \Psi_T \rangle}
.
\label{PIMCopexpdetailE0}
\end{equation}

We conclude that $\tau=0.1$ \textrm{MeV}$^{-1}$
is sufficient to remove excited states and project out the ground state,
by considering the energy gap to the excited states of about $20$ MeV,
along with the variational bound of
about $-23$ MeV with our trial wave function.
Since most operators will be placed in the middle of the path,
we conclude the total imaginary time needs to be about
$\tau=0.2$ \textrm{MeV}$^{-1}$ for convergence.

In Fig. 1 we show the $\alpha$ particle energy versus time step for each potentials studied.
Figure 1 combines Fig. 2, 3, and 4 together, so it is visually easy to see the differences among the ground-state energies predicted by AV6$'$, local N$^2$LO, and AV8$'$ interactions.

\begin{figure}[tbhp!]
\includegraphics[width=8.6cm]{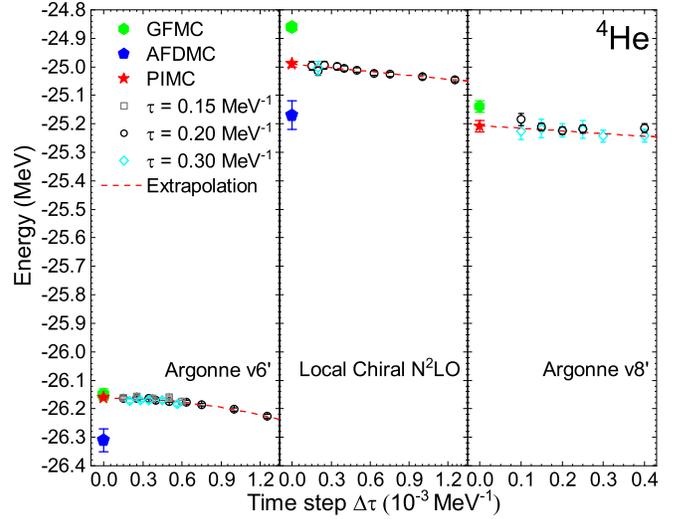}
\caption{Energy vs. time step $\Delta \tau$ for different total imaginary time $\tau$ for $^4$He based on AV6$'$, local chiral N$^2$LO, and AV8$'$ interactions. For the extrapolated ground-state energy at $\Delta \tau=0$ \textrm{MeV}$^{-1}$ for GFMC and PIMC, the size of each symbols represents the corresponding error bar.}
\label{energyall}
\end{figure}

\begin{figure}[tbhp!]
\includegraphics[width=8.6cm]{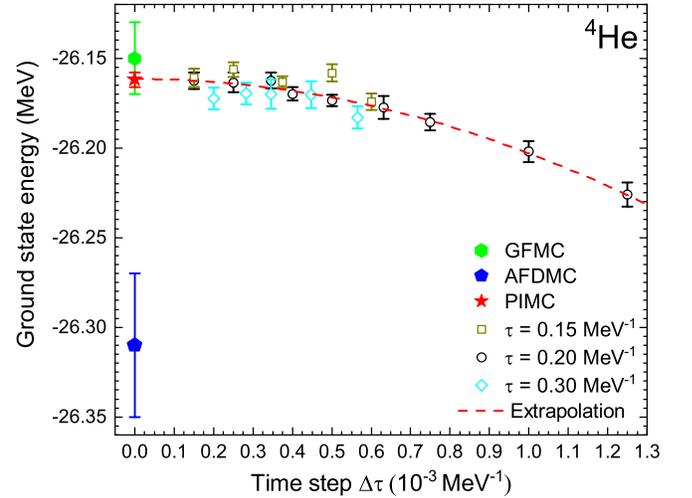}
\caption{Energy vs. time step $\Delta \tau$ for different total imaginary time $\tau$ for $^4$He based on AV6$'$ interaction. The PIMC quadratic extrapolation is based on $\tau=0.2$ \textrm{MeV}$^{-1}$.
The ground-state energy is extrapolated to $-26.162(4)$ \textrm{MeV}.  }
\label{energy}
\end{figure}

\begin{figure}[tbhp!]
\centering
\includegraphics[width=8.6cm]{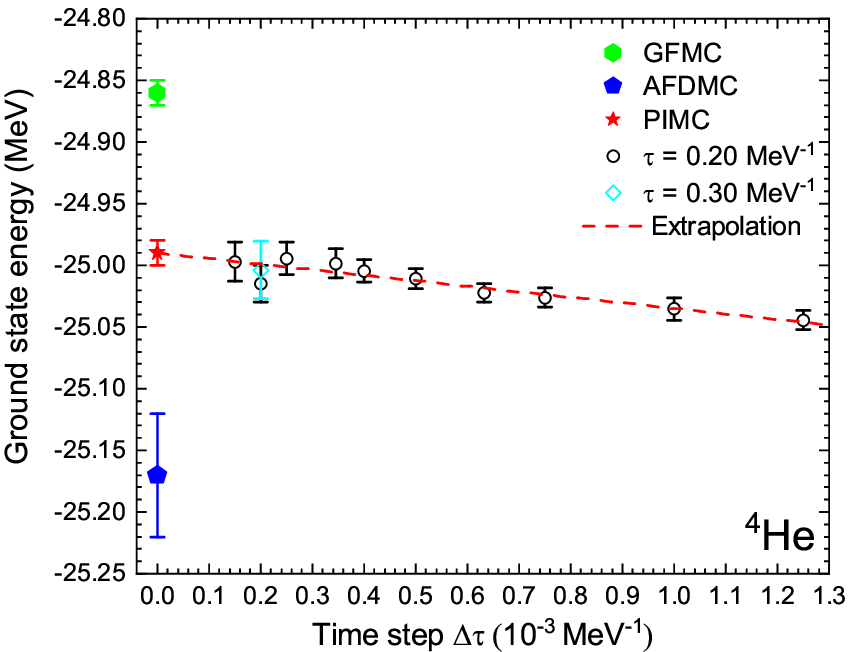}
\caption{Energy vs. time step $\Delta \tau$ for different total imaginary time $\tau$ for $^4$He based on local chiral N$^2$LO interaction. The PIMC linear extrapolation is based on $\tau=0.2$ \textrm{MeV}$^{-1}$. The ground-state energy is extrapolated to $-24.99(1)$ \textrm{MeV}.  }
\label{energyN2LO}
\end{figure}

\begin{figure}[tbhp!]
\centering
\includegraphics[width=8.6cm]{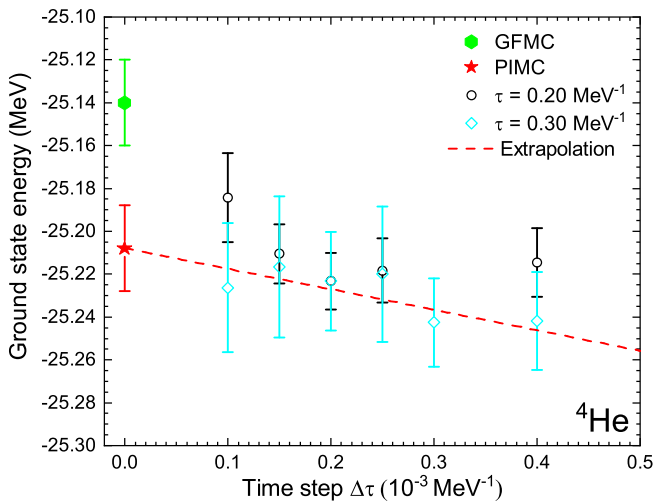}
\caption{Energy vs. time step $\Delta \tau$ for different total imaginary time $\tau$ for $^4$He based on AV8$'$ interaction. The PIMC linear extrapolation is based on $\tau=0.3$ \textrm{MeV}$^{-1}$. The ground-state energy is extrapolated to $-25.21(2)$ \textrm{MeV}.  }
\label{energyV8}
\end{figure}

In Fig. \ref{energy}, we show the AV6$'$ results for the $^4$He ground
state energy $E_0$ for different total imaginary time.
As pointed
out in Eq. (\ref{Oextrapolation}), for AV6$'$,
our PIMC results should
contain $\Delta \tau^2$ order error.
In order to do extrapolation using
Eq.(\ref{Oextrapolation}), we need to find the range of $\Delta \tau$
where the error of $\Delta \tau^2$ dominates.  Roughly , by looking at
Eq.(\ref{Oextrapolation}) we would require $ |\frac{ C_2 \Delta \tau^2
}{C_4 \Delta \tau^4}| \gg 1 $ where $C_4$ is the coefficient for the possible $\Delta \tau^4$ order term. From an analysis of the neglected
commutator terms we
expect $\Delta \tau$ to be around $10^{-4}$ -- $10^{-3} \textrm{ MeV}^{-1}$.
In Fig. \ref{energy} we can observe a clear $\Delta \tau^2$ dependence
when $\Delta \tau < 1.3 \times 10^{-3} \textrm{ MeV}^{-1}$, and by
using Eq.(\ref{Oextrapolation}), the true ground-state energy $E_0$ is
extrapolated to $-26.16(1)$ MeV. This is consistent with the GFMC result
\cite{WiringaPieper02a} of $E_0=-26.15(2)$ MeV, as expected.

In Fig. \ref{energyN2LO} we show the results of $^4$He based on the local
chiral N$^2$LO two-body interaction with a coordinate space cutoff
\cite{Gerzerlis14a,Lynn14a} of $R_0=1.2$ fm.
As discussed in Sec. \ref{propagator} and
Eq.(\ref{ON2LOextrapolation}), due to our approximations in
handling the spin-orbit
operator, the lowest-order time-step error will be order $\Delta \tau$.
In the same
$\Delta \tau < 1.3      \times      10^{-3}$ \textrm{MeV}$^{-1}$
range as AV6', we do observe the $\Delta \tau$ error dominates as
indicated in Eq.(\ref{ON2LOextrapolation}) and the extrapolated ground-state energy is $E_0=-24.99(1)$ MeV.

In Fig. \ref{energyV8} the results of $^4$He based on AV8$'$ interaction are presented.
Due to our approximations in
handling the spin-orbit terms, as in the local
chiral N$^2$LO interaction case, the time-step error is order $\Delta \tau$ as shown in Eq.(\ref{ON2LOextrapolation}).
The linear extrapolation range of the time step is found to be
$\Delta \tau < 0.4    \times      10^{-3}$ \textrm{MeV}$^{-1}$,
and the extrapolated ground-state energy is $E_0=-25.21(2)$ MeV.

\begin{table}[htbp!]
\caption{The ground-state energy of light nuclei based on the AV6$'$ interaction.}
\label{TablePIMC2}%
\begin{tabular*}   {0.48\textwidth}{@{\extracolsep{\fill}}ccccc}
\toprule
$^A Z$   & $E_0$ (MeV) & $\tau$ (MeV$^{-1}$) & $E^{\textrm{GFMC}}_0$ (MeV)  & $E^{\textrm{EXPT}}_0$ (MeV) \\
\hline
$^2$H  & $-2.24(4)$  & 0.5 & &  -2.22 \\ \hline
$^3$H  & $-7.953(5)$  & 0.4 & $-7.95(1)$ & -8.48 \\ \hline
$^3$He  & $-7.336(6)$  & 0.5 & & -7.72  \\ \hline
$^4$He  & $-26.162(4)$  & 0.2 & $-26.15(2)$ & -28.30  \\
\hline\hline
\end{tabular*}
\end{table}

\begin{table}[tbhp!]
\caption{ The ground-state energy of light nuclei based on the local chiral N$^2$LO interaction.}
\label{TablePIMCN2LO}%
\begin{tabular*}   {0.48\textwidth}{@{\extracolsep{\fill}}lcccc}
\hline
\toprule $^A Z$   & $E_0$ (MeV) & $\tau$ (MeV$^{-1}$) & $E^{\textrm{GFMC}}_0$ (MeV) & $E^{\textrm{AFDMC}}_0$ (MeV)  \\ \hline
$^2$H  & $-2.20(1)$  & 0.5 & -2.20 \\ \hline
$^3$H  & $-7.74(1)$  & 0.4 & $-7.74(1)$ & $-7.76(3)$  \\ \hline
$^3$He  & $-7.11(1)$  & 0.5 & $-7.01(1)$ & $-7.12(3)$  \\ \hline
$^4$He  & $-24.99(1)$  & 0.2 & $-24.86(1)$ & $-25.17(5)$ \\
\hline\hline
\end{tabular*}
\end{table}

\begin{table}[tbhp!]
\caption{The ground-state energy of light nuclei based on the AV8$'$ interaction.}
\label{TablePIMCV8'}%
\begin{tabular*}   {0.48\textwidth}{@{\extracolsep{\fill}}lcccc}
\hline
\toprule $^A Z$   & $E_0$ (MeV) & $\tau$ (MeV$^{-1}$) & $E^{\textrm{GFMC}}_0$ (MeV) & $E^{\textrm{AFDMC}}_0$ (MeV)  \\ \hline
$^2$H  & $-2.24(1)$  & 0.5 & \\ \hline
$^3$H  & $-7.79(2)$  & 0.4 & $-7.76(1)$ &  \\ \hline
$^3$He  & $-7.17(4)$  & 0.5 &  &  \\ \hline
$^4$He  & $-25.21(2)$  & 0.3 & $-25.14(2)$ &  \\
\hline\hline
\end{tabular*}
\end{table}

The ground-state energy for all the $A\leq 4$ nuclei are listed in Table \ref{TablePIMC2}
for the AV6$'$ interaction\footnote{In all tables we leave the entry blank if there is no data available.}
,
in Table \ref{TablePIMCN2LO} for the chiral N$^2$LO interaction,
and
in Table \ref{TablePIMCV8'} for AV8$'$ interaction.
The GFMC and AFDMC results \cite{Lynn14a,Lonardoni2018PRC} are listed for comparison.
Overall, all the PIMC, GFMC, and AFDMC results are consistent
with each other within $1\%$ error.
We also listed
$E^{\textrm{EXPT}}_0$ as the experimental value of the binding
energies \cite{WiringaPieper02a,Lee2020FiP,wiki:NBE} for comparison.
The difference between the experimental values and
our results is mainly due to the absence of the three-body interactions
in our calculations.

To confirm our conclusion that
$\tau=0.2$ \textrm{MeV}$^{-1}$
for AV6$'$
and the chiral N$^2$LO,
as well as $\tau=0.3$ \textrm{MeV}$^{-1}$ for AV8$'$ interactions
are
sufficient to project to the ground state,
we calculate the potential energy $V=V^\textrm{NN}+V^\textrm{EM}$ along all the beads
on the path.
The calculation is done in the same way as in Sec. \ref{compalgorithm}.
We use the potential operator $V$ as the operator $\hat{O}$,
and we place the $V$ at each imaginary time position from bead 0 to the last bead $N$ on the path to calculate $\langle {V}(\tau_1) \rangle$,
\begin{eqnarray}
\langle {V}(\tau_1) \rangle
= \frac{ Re  \langle \Psi_T| e^{-H \tau_1}  V e^{-H (\tau-\tau_1) } | \Psi_T \rangle  }
{ Re \langle \Psi_T| e^{-H \tau} | \Psi_T \rangle} ,
\label{potentialbeads}
\end{eqnarray}
where $\tau_1$ ranges from 0 MeV$^{-1}$ to $\tau$ as we put the $V$
operator from bead 0 to bead $N$.  $\langle {V}(\tau_1) \rangle$ is
symmetric around $\tau_1=\tau/2$.  When $\tau/2$ is big enough to project
out the ground state, $\langle {V}(\tau/2) \rangle$ is the ground-state
potential energy.

\begin{figure}[tbhp!]
\includegraphics[width=8.6cm]{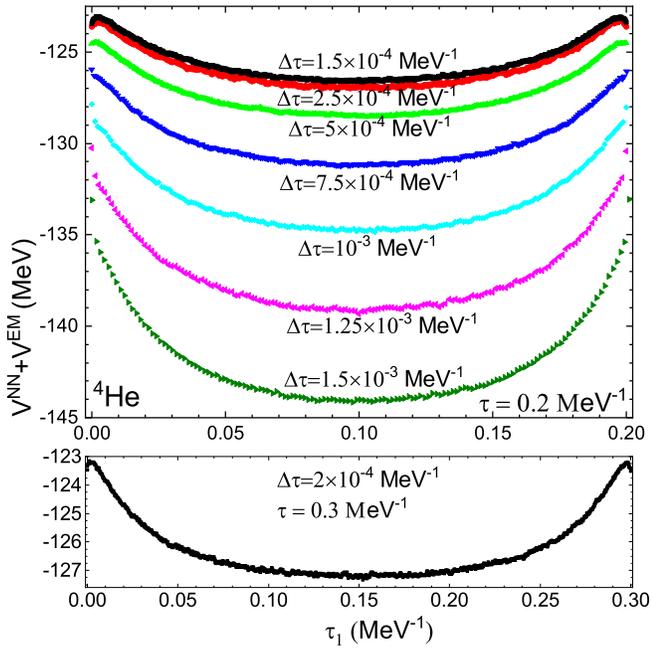}
\caption{The potential energy along the beads for $\tau=0.2$ \textrm{MeV}$^{-1}$ and $\tau=0.3$ \textrm{MeV}$^{-1}$ for $^4$He based on the AV6$'$ interaction. The error bar of the potential energy at each point is not plotted since it is smaller than the size of the corresponding symbol. }
\label{potential}
\end{figure}

\begin{figure}[tbhp!]
\centering
\includegraphics[width=8.6cm]{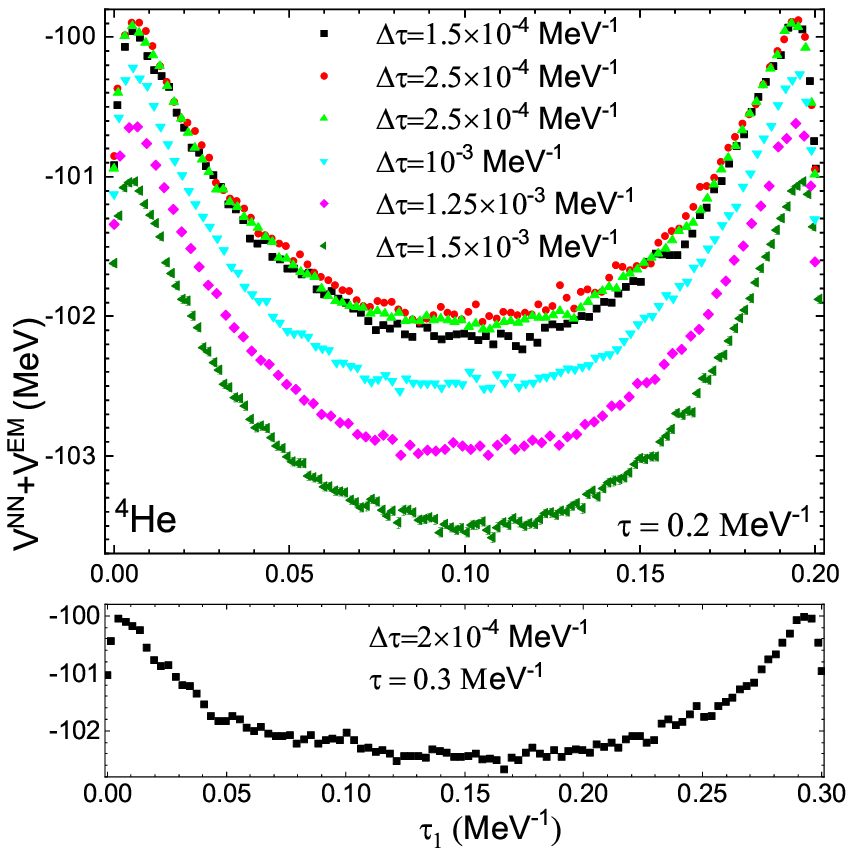}
\caption{The potential energy along the beads for $\tau=0.2$ and 0.3 $\textrm{MeV}^{-1}$ for $^4$He based on the local chiral N$^2$LO interaction. From top to bottom, the $\Delta \tau$ corresponding to each of the plot is displayed. The error bar of the potential energy at each point is equal to or smaller than the size of the corresponding symbol.}
\label{potentialN2LO}
\end{figure}

\begin{figure}[tbhp!]
\centering
\includegraphics[width=8.6cm]{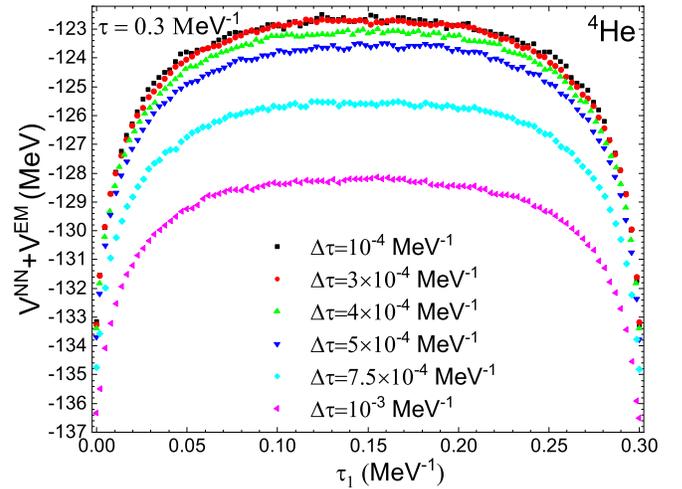}
\caption{The potential energy along the beads for $\tau=0.3$ \textrm{MeV}$^{-1}$ for $^4$He based on the AV8$'$ interaction. From top to bottom, the $\Delta \tau$ corresponding to each of the plot is displayed. The error bar of the potential energy at each point is equal to or smaller than the size of the corresponding symbol.}
\label{potentialV8}
\end{figure}

In Fig. \ref{potential},
in the upper panel
we show the results for $\langle {V}(\tau_1)
\rangle$ as a function of $\tau_1$ for $\tau=0.2$ MeV$^{-1}$. The different
curves are calculations with different time steps, that is, different
numbers of beads.
We see the result converges within error bars when
$\Delta \tau \leq 2.5\times 10^{-4}$ MeV$^{-1}$.
Since the potential is more sensitive to first-order errors in the
wave function, we
verify that the path is long enough for convergence to the
ground state by calculating with
a total $\tau=0.3$ MeV$^{-1}$ in the lower panel. We see that the central
region is essentially flat and in agreement with the $\tau=0.2$ MeV$^{-1}$
central points.

Figure \ref{potentialN2LO} is a similar graph for the local
chiral interaction which also indicates $\tau=0.2$ MeV$^{-1}$
is sufficient.
Figure \ref{potentialV8} shows $\tau=0.3$ \textrm{MeV}$^{-1}$ for AV8$'$ interactions is sufficient.
\footnote{ In order to save computation time, we
did not include spin-orbit interaction in the plot, but it does not alter
the conclusion. Figure \ref{potentialN2LO} is not as smooth as
Fig. \ref{potential} because we ran the calculation for less time
with somewhat higher statistical errors.}

\subsection{Root-mean-square radii}
\label{SECrms}

The  RMS radius $r_m$ is the square root of the
expectation value of the operator
\begin{eqnarray}
 \widehat{r_m^2} = \frac{1}{A} \sum_{i=1}^{A} \left| \hat{\bm{r}}_i - \frac{1}{A}\sum_{j=1}^A \hat{\bm{r}}_j \right|^2 .
\end{eqnarray}
We similarly
define these RMS radii separately for the protons and neutrons by including
isospin projection operators.
Since the RMS radii operators do not commute with $H$,
they require forward walking to calculate them
using diffusion-based methods.
In PIMC, they can be calculated directly.

\begin{table}[tbh!]
\caption{ PIMC results for RMS radii of light nuclei based on AV6$'$ interaction.}
\label{TablePIMC3}%
\begin{tabular*}   {0.48\textwidth}{@{\extracolsep{\fill}}cccc}
\hline
\toprule $^A Z$   & $r_m$ (fm) & $r_p$ (fm) & $r_n$ (fm)  \\ \hline
$^2$H  & $1.98(2)$  &  & \\ \hline
$^3$H  & $1.7261(5)$  & 1.6240(4) & $1.7751(5)$ \\ \hline
$^3$He  & $1.7426(6)$  & 1.7962(6) &  1.6289(5) \\ \hline
$^4$He  & $1.4716(2)$  & 1.4736(2)  & 1.4693(2)  \\
\hline\hline
\end{tabular*}
\end{table}

\begin{table}[tbh!]
\caption{ PIMC results for RMS radii of light nuclei based on local chiral N$^2$LO interaction.}
\label{TablePIMC4}%
\begin{tabular*}   {0.48\textwidth}{@{\extracolsep{\fill}}cccc}
\hline
\toprule $^A Z$   & $r_m$ (fm) & $r_p$ (fm) & $r_n$ (fm)  \\ \hline
$^2$H  & $1.991(1)$  &  & \\ \hline
$^3$H  & $1.7497(7)$  & 1.6430(6) & $1.8007(7)$ \\ \hline
$^3$He  & $1.7725(8)$  & 1.8292(9) &  1.6520(7) \\ \hline
$^4$He  & $1.4860(4)$  & 1.4882(4)  &  1.4834(4)   \\
\hline\hline
\end{tabular*}
\end{table}

\begin{table}[tbh!]
\caption{ PIMC results for RMS radii of light nuclei based on AV8$'$ interaction.}
\label{TablePIMC5}%
\begin{tabular*}   {0.48\textwidth}{@{\extracolsep{\fill}}cccc}
\hline
\toprule $^A Z$   & $r_m$ (fm) & $r_p$ (fm) & $r_n$ (fm)  \\ \hline
$^2$H  & $1.965(3)$  &  & \\ \hline
$^3$H  & $1.760(1)$  & 1.653(1) & $1.811(1)$ \\ \hline
$^3$He  & $1.784(2)$  & 1.841(2) &  1.665(2) \\ \hline
$^4$He  & $1.486(1)$  & 1.488(1)  &  1.483(1)   \\
\hline\hline
\end{tabular*}
\end{table}

In Tables \ref{TablePIMC3}, \ref{TablePIMC4}, and \ref{TablePIMC5},
we list PIMC ground-state nucleon RMS radii based on
the AV6$'$, N$^2$LO local chiral, and AV8$'$ interactions.
For $^2$H, since there is no Coulomb interaction, $r_m=r_p=r_n$.
For other nuclei, however, due to the isospin dependent NN interaction part in the NN interaction)
and the Coulomb interaction, $r_m$, $r_p$ and $r_n$ are, of course,
not the same.
Our results are in
agreement with those in Ref. \cite{Lynn14a}.

\subsection{Density distribution}
\label{secDensity}

The single-particle number density $\rho (r)$ gives
the probability density for one particle to be at distance $r$ from the nuclei's center of mass \cite{Lynn2019ann}.
We normalize so that $\int 4 \pi r^2 \rho (r) dr = 1$, with
\begin{equation}
\hat{\rho}(r) = \frac{1}{A 4\pi r^2} \sum_{i=1}^{A} \delta \left( r - \left| \hat{\bm{r}}_i - \frac{1}{A}\sum_{j=1}^A \hat{\bm{r}}_j  \right|  \right).
\label{rhoravedef}
\end{equation}

\begin{figure}[tbhp!]
\centering
\includegraphics[width=8.6cm]{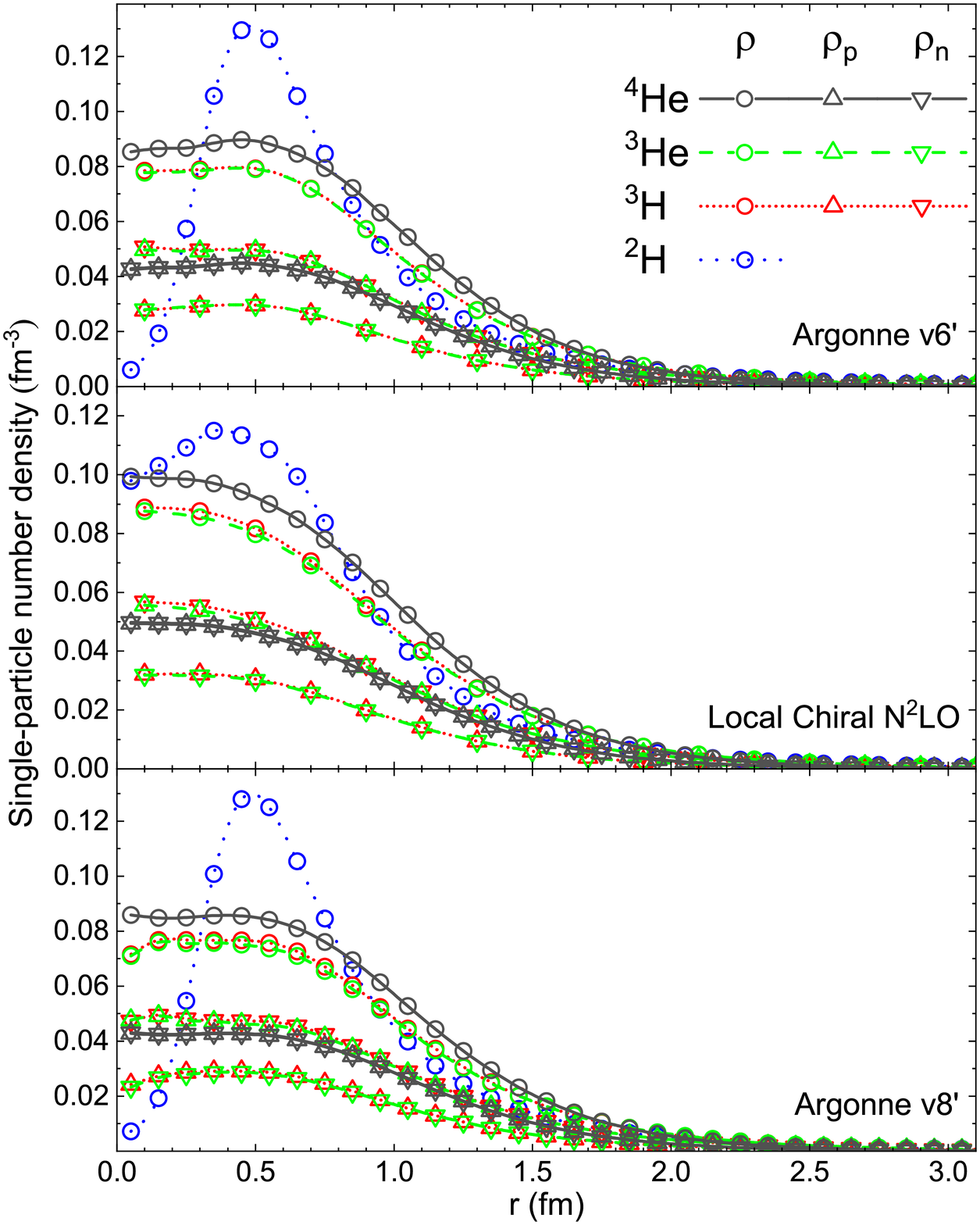}
\caption{The ground-state one-particle number density distributions of $A\leq 4$ light nuclei based on the AV6$'$, local chiral N$^2$LO, and AV8$'$ interactions. The lines are a guide to the eye. $\rho_p$ and $\rho_n$ for deuteron are equal to $\rho/2$. The error bar of one-particle number density at each point is not plotted since it is much smaller than the size of the corresponding symbol.}
\label{rhoall}
\end{figure}

We further define the corresponding proton and neutron
operators as
\begin{equation}
\hat{\rho}_{p(n)}(r) \!=\!  \frac{1}{A4\pi r^2} \sum_{i=1}^{A}
\delta \! \left( r - \left| \hat{\bm{r}}_i - \frac{1}{A}\sum_{j=1}^A
\hat{\bm{r}}_j  \right|   \right) \! P^i_{p(n)},
\end{equation}
where
$P^i_{p}=\frac{1+\tau_{iz}}{2}$ and $P^i_{n}=\frac{1-\tau_{iz}}{2}$ are
the proton and neutron projection operator for nucleon with label $i$.
Since $P^i_{p}+P^i_{n}=1$, we have $ \hat{\rho}(r) = \hat{\rho}_{p}(r)
+ \hat{\rho}_{n}(r)$.  The number density
operators do not commute with $H$, but again, in PIMC, they can be
calculated directly.

In Fig. \ref{rhoall} all the density distributions for $A \leq 4$
light nuclei are presented.

\subsection{Response functions}
\label{SECresponse}

Euclidian response functions\cite{Carlson92a, Carlson94a,
Carlson02a,Lovato13a,Lovato15a} which, in principle, can be
analytically continued to real time, open up the possibility of
exploring the effects of scattering and other interactions with nuclei.
Again, the related operators do
not commute with $H$. So using PIMC is a natural method
to use to calculate response functions.
In an electron-nucleus scattering experiment,
the response of a weakly coupled external probe can be written as the
dynamic structure factor response function $S(k,\omega)$
which can be expanded in the energy
eigenstates \cite{Carlson94a}, as
\begin{equation}
S(k,\omega )
=
\frac{\sum\limits_n \langle \Phi_0 | \rho^\dagger(\bm{k}) | \Phi_n \rangle \langle \Phi_n | \rho (\bm{k})| \Phi_0 \rangle  \delta (\omega+E_0-E_n)  }{\langle \Phi_0 | \Phi_0 \rangle} ,
\label{Skwdef}
\end{equation}
where $\bm{k}$ is the momentum transfer between the final and initial momentum of the nucleus, $\omega$ is the energy transfer between the final and initial energy of the nucleus,
$(k,\omega)$ is the four-momentum carried by the virtual photon \cite{Bacca2014JPG} which is exchanged between the electron and the nucleus,
$E_n$ is eigenenergy of the excited states $|\Phi_n \rangle$, and the $\rho(\bm{k})$ is the coupling operator.

The response function $S(k,\omega)$ is useful because it is related with the scattering cross section and therefore directly connects theory and experiment.
For different scattering processes the couplings of the probe to the nucleus give different $\rho(\bm{k})$ operators.
Here we calculate the Euclidean response function $E(k,\tau)$
which
is related by $S(k,\omega )$ by the Laplace transform,
\begin{eqnarray}
E(k,\tau ) &=&
\displaystyle \int_{0}^{\infty} e^{-\tau (\omega-\omega_{qe})} S(k,\omega ) d\omega \nonumber \\
&=&\frac{ e^{\omega_{qe}\tau } \langle \Psi_T | e^{-H \tau_1}
[\rho^\dagger(\bm{k}) e^{-H\tau } \rho (\bm{k})]
e^{-H \tau_1} | \Psi_T \rangle}
{\langle \Psi_T | e^{-H \tau_1} e^{-H\tau } e^{-H \tau_1} | \Psi_T \rangle},
\nonumber\\
\label{Ektauresponse}
\end{eqnarray}
where $\omega_{qe}=k^2/2m$ and $\tau_1$ is chosen to be large enough to project out the ground state $\Phi_0$ from the trial wave function $\Psi_T$.

The problem of analytically continuing from
$S(k,\omega)$ from $E(k,\tau)$ by inverting the Laplace transform is numerically
unstable, but various methods have made progress \cite{GShen2012PRC,Lovato15a}.

In diffusion QMC such as GFMC and AFDMC, it is often $\langle \Psi_T
|\rho^\dagger(\bm{k}) e^{-H\tau} \rho(\bm{k}) |\Phi_0 \rangle$  that
is calculated.  Since the operator $\rho^\dagger(\bm{k}) e^{-H\tau}
\rho(\bm{k})$ does not commute with the Hamiltonian, $\langle \Psi_T
|\rho^\dagger(\bm{k}) e^{-H\tau} \rho(\bm{k}) |\Phi_0 \rangle$ is a
mixed estimator (although forward walking can improve this) instead of
the true ground-state estimator $\langle \Phi_0 |\rho^\dagger(\bm{k})
e^{-H\tau} \rho(\bm{k}) |\Phi_0 \rangle$. Using PIMC, the response function
calculation is straightforward.

We calculate the $^4$He Euclidean response functions that correspond to
several single-nucleon couplings of $\rho (\bm{k})$ which are similar
with those calculated in Ref. \cite{Carlson94a}. These include the nucleon coupling
$\rho_N (\bm{k})$ , proton coupling $\rho_p (\bm{k})$, isovector
coupling $\rho_\tau (\bm{k})$, spin-longitudinal coupling $\rho_{\sigma
\tau L} (\bm{k})$ and spin-transverse coupling $\rho_{\sigma \tau T}
(\bm{k})$. They are defined as
\begin{eqnarray}
\rho_N (\bm{k}) &=& \sum_{i=1}^{A} e^{i \bm{k}      \cdot \bm{r}_i } , \\
\rho_p (\bm{k}) &=& \sum_{i=1}^{A} e^{i \bm{k}      \cdot \bm{r}_i } \frac{1+\tau_{iz}}{2} ,  \\
\rho_\tau (\bm{k}) &=& \sum_{i=1}^{A} e^{i \bm{k}      \cdot \bm{r}_i } \tau_{iz} , \\
\rho_{\sigma \tau L} (\bm{k}) &=& \sum_{i=1}^{A} e^{i \bm{k}      \cdot \bm{r}_i } (\bm{\sigma }_i      \cdot \hat{\bm{k}}) \tau_{iz} , \\
\rho_{\sigma \tau T} (\bm{k}) &=& \sum_{i=1}^{A} e^{i \bm{k}      \cdot \bm{r}_i } (\bm{\sigma }_i      \times      \hat{\bm{k}}) \tau_{iz} ,
%\label{Accptequationimp}
\end{eqnarray}
where $\bm{r}_i$ is the position of particle $i$.
Since the ground state of $^4$He has total isospin and its $z$ component both 0,
and it is a spherically symmetric object,
the response functions do not depend on the direction of $\bm{k}$.
When calculating the $E(k,\tau)$, the momentum $\bm{k}$ can be averaged
over all the directions.

\begin{figure}[bthp!]
\includegraphics[width=8.6cm]{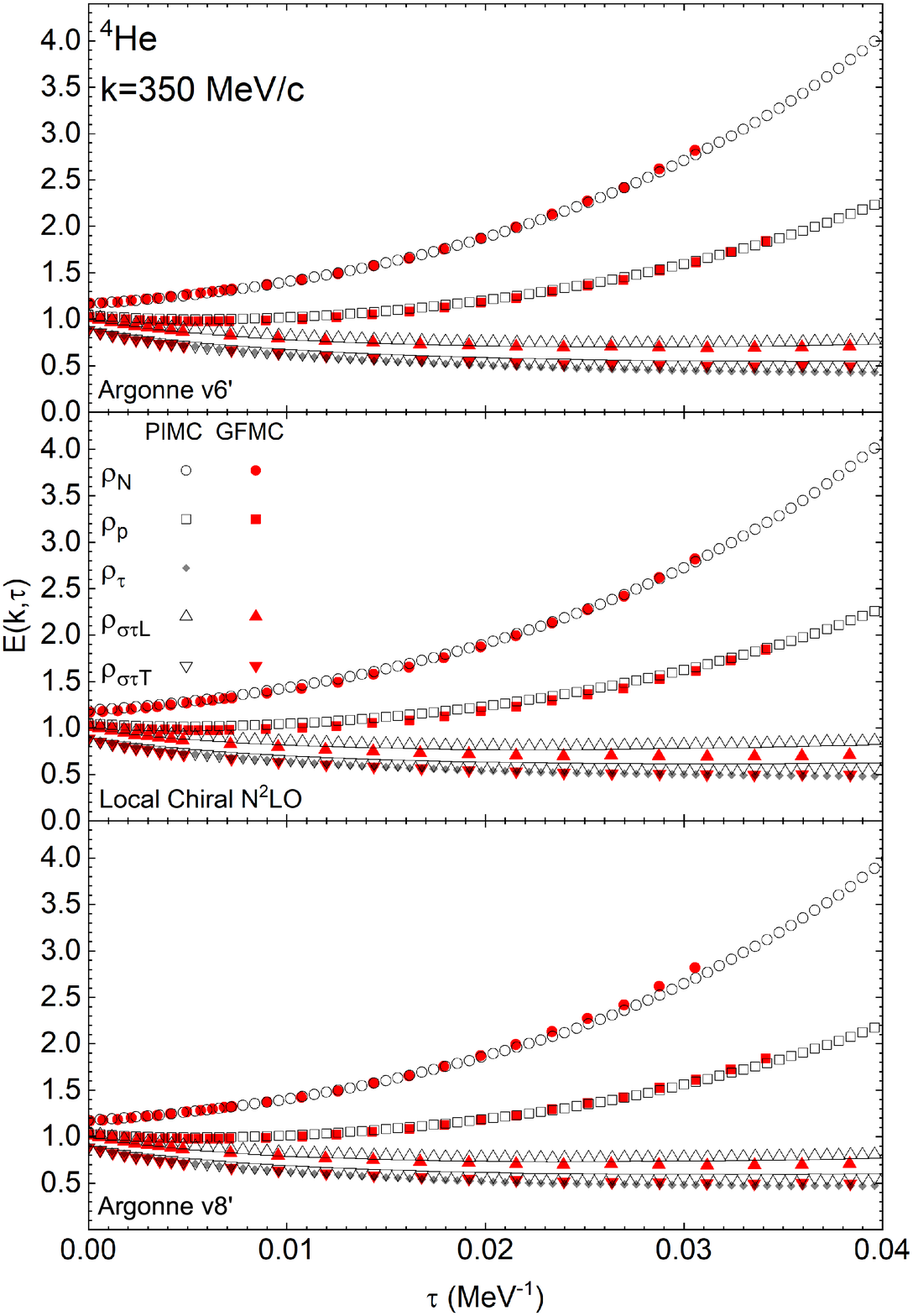}
\caption{Angle averaged response functions $E_\alpha (k,\tau )$ of $^4$He at $k=350$ $\textrm{MeV/c}$ based for AV6$'$, local chiral N$^2$LO, and AV8$'$ interactions. GFMC results \cite{Carlson94a} is based on the AV8$'$ plus the Urbana model-\uppercase\expandafter{\romannumeral8} interactions. The error bar of $E_\alpha (k,\tau )$ at each point is not plotted since it is smaller than the size of the corresponding symbol.}
\label{responseall}
\end{figure}

Figure \ref{responseall} for AV6$'$, local chiral N$^2$LO, and AV8$'$ interactions
shows all of the angle-averaged Euclidean response functions $E_\alpha$ for the
corresponding $\rho_\alpha$ defined above.
Each of $E_\alpha$ has been normalized such that $\lim_{k     \rightarrow \infty} E_\alpha (k,\tau=0)=1$.
The red dots are digitized from the Fig.3 in Ref. \cite{Carlson94a} based on GFMC calculations, so that we can compare our results with theirs.
Clearly our results are comparable with theirs.
The small differences between ours and theirs comes from several sources.
First, we use PIMC which require no forward walking, they use GFMC which requires forward walking.
Second, the interactions are not exactly the same, they used the Argonne $v8$ interaction combined with the Urbana model-\uppercase\expandafter{\romannumeral8}.
Third, we use $\tau_{iz}$ in stead of $\tau_+(i)$ for the isovector, spin-longitudinal, and spin-transverse couplings; this is suggested in their paper when dealing with an isoscalar target like the ground state $^4$He and with interactions that conserve the number of protons and neutrons.

\subsection{Computational scaling}
Overall, for PIMC, the scaling of required computational time with the number of nucleons and the number of time slices are comparable with those for GFMC with a somewhat larger prefactor.
PIMC calculations for propagating a short time step are the same as GFMC
and scale with the number of spin/isospin states
$ N_{\rm{tot}}= \frac{A!}{Z!(A-Z)!} 2^A$
if isospin breaking terms are included in the propagation.

The path integral requires additional updates of the whole path when new
path positions are included which are absent in GFMC. We expect these to
scale with the path length. But typically
require less than an order of magnitude
of additional computations for the sampling methods we have used.

In order to take advantage of the ability to calculate operator expections
in the middle of the path,
PIMC typically needs twice the total imaginary time required to converge the energy.

An advantage of the path-integral method is that because of the Metropolis
sampling, the time-step errors even for the simplest short-time propagator
can be readily controlled.
We can extrapolate to the zero time-step
limit using larger time steps than GFMC.
That is, in a typical GFMC
calculation, an approximate propagator is sampled, and the ratio of
the correct propagator at the sampled position and the approximation
is included in a weight. In the path integral version instead of this
weight, the Metropolis accept/reject step is used. This replaces the
fluctuating weight, with its attendant increase in variance with
Metropolis rejections. In our formalism, rejections are inexpensive,
with the result that we find much lower variance for larger time
steps, and can therefore take larger time steps and still extrapolate
to the zero time-step limit.

The calculation here use very simple trial wave functions in order to show the convergence for a variety of expectation values and responses. Much better trial functions are typically used in GFMC calculations. These will improve our convergence and lower variances.

GFMC calculations typically use a pair-product propagator. Its use in the path integral calculation is straightforward and it would be used for heavier nuclei. We expect that this would allow even larger time steps and more efficient calculations.

Overall, by using larger steps, better trial wave functions, pair-product propagator, we expect the PIMC scaling to be similar to GFMC, with total
calculation times
essentially proportional to $4^A$.
\section{Summary and Outlook}
\label{summary}

By using a variety of sampling techniques
along with
the optimized strategies of calculating and updating the path
we successfully performed real-space accurate ground-state nuclear PIMC calculation of light nuclei, based on local chiral with N$^2$LO, AV6$'$,
and AV8$'$ interactions.
From our analysis in Sec. \ref{secGS} we can conclude that our choices of total
imaginary time indeed projected out the ground states.
Also, the extrapolation behavior between
$\langle \hat{O} \rangle$ and $\langle \hat{O}(\Delta \tau) \rangle$ are as expected.
As discussed in Sec. \ref{secErrEst}, we
verified the expected path discretization errors
and extrapolated the results to zero time-step error.
These results show that the nuclear PIMC method is accurate and reliable.

Our PIMC ground-state energies of light nuclei are consistent with the results from GFMC and AFDMC as shown in Tables \ref{TablePIMC2}, \ref{TablePIMCN2LO} and \ref{TablePIMCV8'}.
For operators which do not commute with Hamiltonian and cannot be obtained from diffusion Monte Carlo
based methods such GFMC and AFDMC without forward walking, we easily
get reliable results.
We calculated accurate particle number density distributions $\rho (r)$,
RMS radii, and the angle-averaged Euclidean response functions.

These results show the power of real-space
PIMC calculation for light nuclei.
In our PIMC calculation,
we find that even with our simple Trotter
break up of the propagator,
a time step of $10^{-4}$ -- $10^{-3} \textrm{ MeV}^{-1}$ works reasonably well.
Depending on $\tau$ and $\Delta \tau$, the number of beads for PIMC ranges from 41 to 3001.
In the most time consuming case,
such as the AV8$'$ interaction for $\tau=0.3 \textrm{ MeV}^{-1}$ and $\Delta \tau=10^{-4} \textrm{ MeV}^{-1}$, we need 3000 short-time propagators and therefore 3001 beads.
Such a calculation,
to reach a less than $0.02\textrm{ MeV}$ error in the ground-state energy calculation,
takes about 40000 core hours for the
interactions used.
Results with
41 beads only take a few hundred core hours.

Our calculations here are for $A\leq 4$.
For $A>4$, GFMC has a serious
fermion sign/phase problem, and our path integral calculations will as
well.
Initially for such calculations
we expect to be able to use constrained path methods as
in GFMC calculations\cite{CarlsonRMP15a}
or as in finite-temperature path-integral calculations\cite{Ceperley95a}.
These can be either based on the trial wave-function phase as in GFMC
or from the trial wave function at the ends and the propagator as in some
constrained path-integral methods.
One advantage to the path-integral
formalism is that, unlike GFMC, we can have an upper-bound principle for
all of these constraints if we calculate the expectation value of the
Hamiltonian at the center of the path with equivalent constraints on
the left and right paths. This can open up the possibilities of
optimizing both the trial functions and the constraints within the
path-integral method.

Our $^4$He calculations here are a benchmark test for
future PIMC calculations of larger nuclei.
We also believe that, based on
what we have in this paper, a nuclear PIMC finite-temperature calculation
of $\alpha$ particles or neutron matter are feasible.
Three-body interactions can be readily included in a straightforward
manner to perform more
accurate PIMC calculations.

\begin{acknowledgments}
We thank professor Francesco Pederiva for helpful discussions.
This work was supported by the National Science Foundation Grant No. PHY-1404405. We acknowledge Research Computing at Arizona State University for providing HPC and storage resources that have contributed to the research results reported within this paper. All calculations were performed on the Agave research computing cluster.
\end{acknowledgments}

\appendix
\section{Path Calculation}
\label{secCalcPath}

The $A_{lr} (\mathcal{R})$ and $B_{lr} (\mathcal{R})$ in Eq. (\ref{OABP})
are real functions,
\begin{eqnarray}
% \nonumber % Remove numbering (before each equation)   \! can be used to make expressions shorter.
A_{lr} (\mathcal{R}) &=& \frac{Re[ g(\mathcal{R})_{lr}]}{|Re[ f(\mathcal{R})_{lr}]|  } ,
\label{Alrform}
\\
B_{lr} (\mathcal{R}) &=& \frac{Re[ f(\mathcal{R})_{lr}]}{|Re[ f(\mathcal{R})_{lr}]|  },
\label{Blrform}
\end{eqnarray}

$P_{lr} (\mathcal{R})$ in Eq. (\ref{OABP}) is the probability distribution,
\begin{equation}
P_{lr} (\mathcal{R}) =
\frac{|Re[ f(\mathcal{R})_{lr}  ]|}{ \mathcal{N}  },  \label{PlrR}
\end{equation}
which is normalized,
$ \sum_{l}\sum_{r} \int d \mathcal{R} P_{lr}(\mathcal{R}) =1$.
The normalization factor
$ \mathcal{N}=\sum_l \sum_r \int \mathcal{D} \mathcal{R} |Re[ f(\mathcal{R})_{lr}]|$ cancels in the Metropolis algorithm implementation.

Function $f(\mathcal{R})_{lr}$ comes from the denominator of Eq. (\ref{OABP1}) and it does not depend on where the operator $\hat{O}$ is placed.
Function $g(\mathcal{R})_{lr}$ comes from the numerator of Eq. (\ref{OABP1}), and it depends on which bead $M$ the operator $\hat{O}$ is placed at. They can be written as
\begin{eqnarray}
f(\mathcal{R})_{lr} &=&
f^V_{lr}(\mathcal{R})
g^F(\mathcal{R}) , \label{fRlr} \\
g(\mathcal{R})_{lr}  &=&
g^V_{lr,M}(\mathcal{R})
 g^F(\mathcal{R}). \label{gRlr}
\end{eqnarray}
Function $g^F(\mathcal{R})$ in Eqs. (\ref{fRlr})--(\ref{gRlr}) is
\begin{equation}
g^F(\mathcal{R}) = \prod_{I=0}^{N-1} G^f_{I, I\!+\!1} . \label{gF}
\end{equation}
The Gaussian free particle propagator,
 $G^f_{I, I+1}$ is defined in
Eq.(\ref{propagatorFree}) and
connects beads $R_I$ and $R_{I+1}$.
The product $\prod_{I=0}^{N-1}  G^f_{I, I+1}$
in the functions $A_{lr} (\mathcal{R})$ and $B_{lr} (\mathcal{R})$,
cancels. They become
\begin{eqnarray}
% \nonumber % Remove numbering (before each equation)   \! can be used to make expressions shorter.
A_{lr} (\mathcal{R}) &=& \frac{Re[ g^V_{lr,M}(\mathcal{R})]}{|Re[ f^V_{lr}(\mathcal{R})]|  } \label{Alr}, \\
B_{lr} (\mathcal{R}) &=& \frac{Re[ f^V_{lr}(\mathcal{R})]}{|Re[ f^V_{lr}(\mathcal{R})]|  }  \label{Blr}.
\end{eqnarray}

To simplify the notations in the expressions of functions
$g^V_{lr,M}(\mathcal{R})$
and
$f^V_{lr}(\mathcal{R})$,
we define states $| \psi^r_M (\mathcal{R})
\rangle $ and $| \psi^l_M (\mathcal{R})\rangle$ where the Dirac notation
is used for the spin/isospin states, while the position parts of the
wave function are evaluated at the bead positions given by $\mathcal{R}$,
\begin{widetext}
\begin{eqnarray}
% \nonumber % Remove numbering (before each equation)   \! can be used to make expressions shorter.
| \psi^r_M (\mathcal{R}) \rangle  &      \equiv      &
\begin{dcases}
 \sum\limits_{S}
U_V(R_M,\tfrac{\Delta\tau}{2})
U_V^\dagger(R_{M},\tfrac{\Delta\tau}{2})
\mathcal{G}(R_M,R_{M+1})
U_V(R_{M+1},\tfrac{\Delta\tau}{2})
U_V^\dagger(R_{M+1},\tfrac{\Delta\tau}{2})
\mathcal{G}(R_M,R_{M+2})...&
\\
%\hskip .1\columnwidth
\ \ \ \ \ \ ...\
U_V^\dagger(R_{N-1},\tfrac{\Delta\tau}{2})
\mathcal{G}(R_{N-1},R_{N})
U_V(R_{N},\tfrac{\Delta\tau}{2})
| S \rangle \langle R_N  S | \psi^r_T \rangle,
& \hspace{-4.5em} (M      \neq       N,  0)
\\
 \sum\limits_{S}
U_V(R_N,\tfrac{\Delta\tau}{2})
| S \rangle \langle R_N  S | \psi^r_T \rangle,
& \hspace{-4.5em}  (M = N) \\
 \sum\limits_{S} \!
U_V^\dagger(R_{0},\tfrac{\Delta\tau}{2})
\mathcal{G}(R_{0},R_{1})
U_V(R_{1},\tfrac{\Delta\tau}{2})
U_V^\dagger(R_{1},\tfrac{\Delta\tau}{2})
\mathcal{G}(R_{1},R_{2})
U_V(R_{2},\tfrac{\Delta\tau}{2})
U_V^\dagger(R_{2},\tfrac{\Delta\tau}{2})
\mathcal{G}(R_{2},R_{3})
 ... &\\
\ \ \ \ \ \ ...\
U_V^\dagger(R_{N-1},\tfrac{\Delta\tau}{2})
\mathcal{G}(R_{N-1},R_{N})
U_V(R_{N},\tfrac{\Delta\tau}{2})
| S \rangle \langle R_N  S | \psi^r_T \rangle.
& \hspace{-4.5em} (M = 0)
\end{dcases} \label{psirmRdef}
%\end{eqnarray}
%\begin{eqnarray}
\\
| \psi^l_M (\mathcal{R})\rangle  &      \equiv      &
\begin{dcases}
\sum\limits_{S}
U_V(R_M,\tfrac{\Delta\tau}{2})
U_V^\dagger(R_{M},\tfrac{\Delta\tau}{2})
\mathcal{G}^*(R_{M},R_{M-1})
U_V(R_{M-1},\tfrac{\Delta\tau}{2})
U_V^\dagger(R_{M-1},\tfrac{\Delta\tau}{2})
\mathcal{G}^*(R_{M-1},R_{M-2})
...&\\
%\hskip .1\columnwidth
\ \ \ \ \ \ ...\
\mathcal{G}^*(R_{2},R_{1})
U_V(R_{1},\tfrac{\Delta\tau}{2})
U_V^\dagger(R_{1},\tfrac{\Delta\tau}{2})
\mathcal{G}^*(R_{1},R_{0})
U_V(R_{0},\tfrac{\Delta\tau}{2})
| S \rangle \langle R_0 S | \psi^l_T \rangle,
& \hspace{-4.5em}  (M     \neq      0,N) \\
\sum\limits_{S}
U_V(R_{0},\tfrac{\Delta\tau}{2})
| S \rangle \langle R_0  S | \psi^l_T \rangle,
& \hspace{-4.5em}  (M=0) \\
\sum\limits_{S}\!
U_V^\dagger(R_{N},\tfrac{\Delta\tau}{2})
\mathcal{G}^*(R_{N},R_{N-1})
U_V(R_{N-1},\tfrac{\Delta\tau}{2})
U_V^\dagger(R_{N-1},\tfrac{\Delta\tau}{2})
\mathcal{G}^*(R_{N-1},R_{N-2})
U_V(R_{N-2},\tfrac{\Delta\tau}{2})
...&\\
%\hskip .1\columnwidth
\ \ \ \ \ \ ...\
\mathcal{G}^*(R_{2},R_{1})
U_V(R_{1},\tfrac{\Delta\tau}{2})
U_V^\dagger(R_{1},\tfrac{\Delta\tau}{2})
\mathcal{G}^*(R_{1},R_{0})
U_V(R_{0},\tfrac{\Delta\tau}{2})
| S \rangle \langle R_0 S | \psi^l_T \rangle.
& \hspace{-4.5em}  (M=N)
\end{dcases} \label{psilmRdef}
%\hspace{-5em}
\end{eqnarray}
\end{widetext}
Note that for AV6$'$ which does not have spin-orbit term, the $\mathcal{G}(R,R')$ factors are 1.

Both $| \psi^l_M (\!\mathcal{R}\!)\rangle$
and $| \psi^r_M (\!\mathcal{R}\!)\rangle$
can be written in the $A$-particle spin-isospin basis
$| S \rangle$ such that
%\begin{widetext}
\begin{align}
| \psi^r_M (\!\mathcal{R}\!) \rangle
&= \sum_S | S \rangle \langle S | \psi^r_M (\!\mathcal{R}\!) \rangle
= \sum_S \phi^r_{M}(\mathcal{R},S) | S \rangle , \label{phirMexpansion}
\\
| \psi^l_M (\!\mathcal{R}\!)\rangle
&= \sum_S | S \rangle \langle S | \psi^l_M (\!\mathcal{R}\!)\rangle
= \sum_S  \phi^l_{M}(\mathcal{R},S) | S \rangle,
\label{philMexpansion}
\end{align}
%\end{widetext}
where $\phi^r_{M}(\mathcal{R},S)$ and $\phi^l_{M}(\mathcal{R},S)$ are the  corresponding probability amplitude.

The function $g^V_{lr,M}(\mathcal{R})$ depends on which bead, $M$,
the operator $\hat{O}$ is placed.
For a given $M$, with the defined
$| \psi^r_M (\!\mathcal{R}\!) \rangle$
and
$| \psi^l_M (\!\mathcal{R}\!)\rangle $,
it is
\begin{widetext}
\begin{equation}
g^V_{lr,M}(\mathcal{R}) =
\begin{dcases}
\langle \psi^l_{M\!-\!1} (\mathcal{R}) |
\mathcal{G}(R_{M-1},R_{M})
U_V(R_{M},\tfrac{\Delta\tau}{2})
\hat{O}
U_V^\dagger(R_{M},\tfrac{\Delta\tau}{2})
\mathcal{G}(R_{M},R_{M+1})
| \psi^r_{M\!+\!1} (\mathcal{R}) \rangle,
 &  (M      \neq      0, N) \\
\sum\limits_{S} \langle \psi^l_T| \hat{O} | R_0 S\rangle \langle S | \psi^r_0 (\mathcal{R}) \rangle,
 &  (M=0) \\
\sum\limits_{S} \langle \psi^l_N (\mathcal{R}) | S \rangle \langle R_N S| \hat{O} | \psi^r_T \rangle.
 & (M=N)
\end{dcases}
\label{gVRlrdef2}
\end{equation}
\end{widetext}

The integer $M$ in $g^V_{lr}(\mathcal{R}) $ is usually set so that
$\hat O$ operates at the central bead.
This guarantees that the calculation gives
the ground-state expectation value of $\hat{O}$
if the total imaginary time $\tau$ is large.

If the operator $\hat{O}$ commutes with Hamiltonian, then it is
often convenient to calculate $\hat{O}$ by operating on the trial wave function.
The corresponding $g(\mathcal{R})_{lr}$ can be calculated as,
\begin{equation}
% \nonumber % Remove numbering (before each equation)   \! can be used to make expressions shorter.
\hspace{-0.5em} g(\mathcal{R})_{lr} =  \frac{
\left[ g^V_{lr,M}(\mathcal{R}) |_{M=0} + g^V_{lr,M}(\mathcal{R}) |_{M=N}  \right] }{2}
g^F(\mathcal{R}) .
\label{gRlrOH}
\end{equation}
This is for example how we calculate the ground-state energy $\hat{O}=H$.

The function $f^V_{lr}(\mathcal{R})$ does not depend on $M$.
Its value is
the same as long as the left and right orders $l$, $r$, and $\mathcal{R}$
are the same.  For any integer $M$ ranges from 0 to $N-1$, it is
calculated as
\begin{align}
% \nonumber % Remove numbering (before each equation)   \! can be used to make expressions shorter.
f^V_{lr}(\mathcal{R})
&= \langle \psi^l_M (\mathcal{R}) | \mathcal{G}(R_{M},R_{M+1}) | \psi^r_{M+1} (\mathcal{R}) \rangle
\nonumber \\
&= \sum_S \langle \psi^l_M (\mathcal{R}) | S \rangle
\langle S| \mathcal{G}(R_{M},R_{M+1}) | \psi^r_{M+1} (\mathcal{R}) \rangle.
\nonumber \\
\label{fVRlrN2LOdef2}
\end{align}

\section{Path Updating}
\label{secUpdatePathStrategy}
Based on Eq.(\ref{psirmRdef}) and Eq.(\ref{psilmRdef}) we write recursion
relations
\begin{widetext}
\begin{eqnarray}
% \nonumber % Remove numbering (before each equation)   \! can be used to make expressions shorter.
| \psi^r_M (\mathcal{R}) \rangle  &=&
\begin{dcases}
\sum\limits_{S}
U_V(R_{M},\tfrac{\Delta\tau}{2})
U_V^\dagger(R_{M},\tfrac{\Delta\tau}{2})
\mathcal{G}(R_{M},R_{M+1})
| S \rangle
\langle S| \psi^r_{M+1} (\mathcal{R}) \rangle,
 & (M \neq  N,  0) \\
\sum\limits_{S}
U_V(R_{N},\tfrac{\Delta\tau}{2})
| S \rangle \langle R_N  S | \psi^r_T \rangle,
&  (M = N) \\
\sum\limits_{S} \!
U_V^\dagger(R_{0},\tfrac{\Delta\tau}{2})
\mathcal{G}(R_{0},R_{1})
| S \rangle \langle S | \psi^r_1 (\mathcal{R}) \rangle,
 &  (M =0)
\end{dcases} \label{phirMRrecur}
%\end{eqnarray}
%\begin{eqnarray}
\\
| \psi^l_M (\mathcal{R})\rangle  &=&
\begin{dcases}
 \sum\limits_{S}
U_V(R_{M},\tfrac{\Delta\tau}{2})
U_V^\dagger(R_{M},\tfrac{\Delta\tau}{2})
\mathcal{G}^*(R_{M},R_{M-1})
| S \rangle
\langle S | \psi^l_{M-1} (\mathcal{R}) \rangle,
 & (M\neq 0,N) \\
 \sum\limits_{S}
U_V(R_{0},\tfrac{\Delta\tau}{2})
| S \rangle \langle R_0  S | \psi^l_T \rangle,
 &  (M=0) \\
 \sum\limits_{S}\!
U_V^\dagger(R_{N},\tfrac{\Delta\tau}{2})
\mathcal{G}^*(R_{N},R_{N-1})
| S \rangle
\langle S | \psi^l_{N-1} (\mathcal{R}) \rangle.
 &  (M=N)
\end{dcases} \label{philMRrecur}
%\hspace{-5em}
\end{eqnarray}
\end{widetext}
We then directly calculate
$| \psi^r_{N-1} (\mathcal{R}) \rangle $ from $| \psi^r_N (\mathcal{R}) \rangle $, then $| \psi^r_{N-2} (\mathcal{R}) \rangle $ from $| \psi^r_{N-1} (\mathcal{R}) \rangle $, \ldots, until $| \psi^r_{M+1} (\mathcal{R}) \rangle $ from $| \psi^r_{M+2} (\mathcal{R}) \rangle $
from right to left. Similarly we calculate all of the $| \psi^l_{M} (\mathcal{R})
\rangle $ from $| \psi^l_{M-1} (\mathcal{R}) \rangle $.

Once we
have the corresponding $| \psi^r_{M} (\mathcal{R}) \rangle $ and $|
\psi^l_{M} (\mathcal{R}) \rangle $, we use Eqs. (\ref{fVRlrN2LOdef2})
and (\ref{gVRlrdef2}) to calculate  $f^V_{lr}(\mathcal{R})$
and $ g^V_{lr,M}(\mathcal{R}) $ and then $A_{lr}(\mathcal{R})$ and
$B_{lr}(\mathcal{R})$ and thus $\langle \hat{O} (\Delta \tau )  \rangle$.
This recursive updating strategy for $| \psi^r_{M} (\mathcal{R}) \rangle
$ and $| \psi^l_{M} (\mathcal{R}) \rangle $ is efficient when
updating part of the path.

\bibliographystyle{apsrev4-2}
%\bibliography{v6pimc1}

\end{document}